\documentclass[11pt]{article}

\usepackage[final]{acl}

\usepackage{times}
\usepackage{latexsym}

\usepackage[T1]{fontenc}

\usepackage[utf8]{inputenc}

\usepackage{microtype}

\usepackage{inconsolata}

\usepackage{graphicx}

\usepackage{amsmath}
\usepackage{amssymb}   
\usepackage{amsfonts}  
\usepackage{booktabs}   
\usepackage{graphicx}   
\usepackage{bbding}
\usepackage{multirow}
\usepackage{subcaption} 
\usepackage{caption}
\usepackage{CJKutf8}
\usepackage{tabularx}
\usepackage{array}

\title{When Tone and Words Disagree: Towards Robust Speech Emotion Recognition under Acoustic-Semantic Conflict}

\author{
 \textbf{Dawei Huang\textsuperscript{1,2}},
 \textbf{Yongjie Lv\textsuperscript{1}},
 \textbf{Ruijie Xiong\textsuperscript{1}},
 \textbf{Chunxiang Jin\textsuperscript{1}}, 
 \textbf{Xiaojiang Peng\textsuperscript{2}\thanks{Corresponding author.}},
\\
 \textsuperscript{1}Inclusion AI, Ant Group,
  \textsuperscript{2} Shenzhen University,
\\
}

\begin{document}
\maketitle
\begin{abstract}
Speech Emotion Recognition (SER) systems often assume congruence between vocal emotion and lexical semantics. 
However, in real-world interactions, acoustic-semantic conflict is common yet overlooked, where the emotion conveyed by tone contradicts the literal meaning of spoken words. 
We show that state-of-the-art SER models, including ASR-based, self-supervised learning (SSL) approaches and Audio Language Models (ALMs), suffer performance degradation under such conflicts due to semantic bias or entangled acoustic–semantic representations. 
To address this, we propose the \textbf{Fusion Acoustic-Semantic (FAS)} framework, which explicitly disentangles acoustic and semantic pathways and bridges them through a lightweight, query-based attention module. 
To enable systematic evaluation, we introduce the \textbf{Conflict in Acoustic-Semantic Emotion} \textbf{(CASE)}, the first dataset dominated by clear and interpretable acoustic-semantic conflicts in varied scenarios.
Extensive experiments demonstrate that FAS consistently outperforms existing methods in both in-domain and zero-shot settings.
Notably, on the CASE benchmark, conventional SER models fail dramatically, while FAS sets a new SOTA with 59.38\% accuracy.
Our code and datasets is available at \href{https://github.com/24DavidHuang/FAS}{https://github.com/24DavidHuang/FAS}.
\end{abstract}

\section{Introduction}

Speech Emotion Recognition (SER), a fundamental task in affective computing, aims to automatically identify the emotional state of a speaker from their vocal expressions.
Current SER methods \cite{emotion2vec, Vesper, CLAP, hubert, whisper, Wavlm} have demonstrated remarkable performance on standard academic benchmarks such as IEMOCAP \cite{IEMOCAP} and MELD \cite{MELD}. However, this success is largely confined to scenarios of acoustic-semantic congruence, where the prosodic cues in speech (acoustics) align with the literal meaning of the spoken content (semantics)—for instance, expressing "What a beautiful day!" in a joyful tone.

The crux of the problem is that emotion is inherently complex and frequently manifests under conditions of acoustic-semantic conflict.
Real-world communication is replete with nuanced expressions like sarcasm, schadenfreude (gloating) or cold fury, where a speaker's true emotion, conveyed through acoustic cues, is decoupled or even antithetical to the semantic content of their utterance. 
For example, upon learning that a colleague has been promoted, someone might say \textit{"Congratulations on your promotion!"} in a flat or resentful tone, subtly revealing underlying envy rather than joy.
In these prevalent yet challenging scenarios, the performance of current SER methods collapses.

This vulnerability is systemic across current SER paradigms. speech-text pre-trained encoders \cite{whisper, CLAP} exhibit a strong semantic bias, causing them to be "poisoned" by the literal meaning of words. 
Self-supervised learning (SSL) methods \cite{hubert, wav2vec2.0, wav2vec, Wavlm} produce entangled representations where affective and semantic information are conflated, making ambiguity difficult to resolve. 
Furthermore, explicit multimodal approaches \cite{MM_Conflict, SZTU-CMU} struggle, as their current modality fusion mechanisms often lack a robust strategy to arbitrate between conflicting signals, defaulting to the misleading modality. 
Recent ALMs, despite their impressive capabilities, depend on LLM-aligned encoders (e.g. Whisper \cite{whisper}, CLAP \cite{CLAP}) that prioritize semantics over prosody, depriving the LLM of affective cues during emotional conflict. 
All these brittleness limit the reliable application of SER models in unconstrained real-world environments.

What's more, current SER evaluation is limited by datasets biases: most benchmarks feature predominantly congruent emotional expressions. 
While in-the-wild datasets (e.g., MELD, IEMOCAP) contain occasional acoustic-semantic conflicts, such cases are sparse and unstructured. Critically, no existing resource provides a high-density, controlled setting to systematically evaluate robustness under emotional conflict—leaving a key aspect of real-world performance unassessed.

To address these challenges, this paper introduces a dual-pronged contribution.
First, we propose an innovative \textbf{Fusion Acoustic-Semantic (FAS)} framework, designed to explicitly disentangle acoustic and semantic information from speech. The FAS uniquely employs an audio tokenizer, inspired by recent advances in Text-to-Speech generation, to extract low-dimensional acoustic tokens, while concurrently utilizing a pre-trained module to capture high-dimensional semantic information.
A lightweight, query-based module is then introduced to integrate disentangled features and make robust predictions.

Second, to validate our proposed framework and to provide a much-needed resource for the community, we released the \textbf{Conflict in Acoustic-Semantic Emotion (CASE)} Benchmark. Unlike conventional datasets, CASE is constructed with a high concentration of logical, interpretable, and scenario-driven conflict samples. It serves not only as a challenging testbed for evaluating model robustness but also as a valuable corpus for researchers to study the interplay between acoustics and semantics in human emotion expression.
The main contributions of this paper are summarized as follows:
\begin{itemize}
    \item We are the first to systematically investigate the problem of affective-semantic conflict in SER, showing that existing methods degrade significantly on such samples.
    \item We introduce the \textbf{CASE} benchmark, the first standardized dataset designed to assess the robustness of SER models against acoustic-semantic conflict.
    \item We propose the \textbf{FAS} framework, designed to resolve emotional ambiguity by disentangling acoustic cues and semantic content. Through extensive experiments, we demonstrate the superiority of our FAS over SOTA baselines on all benchmarks.
\end{itemize}

\section{Related Work}
\label{sec:Related Work}

\subsection{Speech Emotion Recognition}
Recent advances in Speech Emotion Recognition (SER) have been predominantly driven by large-scale pre-trained models like WavLM \cite{Wavlm}, Whisper \cite{whisper} and HuBERT \cite{hubert}.
A significant body of work \cite{emotion2vec, Vesper, MFGCN} focuses on transferring, distilling and adapting the representations from these powerful encoders to SER task. 

However, ASR-based encoders such as Whisper \cite{whisper} and CLAP \cite{CLAP}, while powerful, exhibit a strong semantic bias due to their pre-training objective, making them highly susceptible to being misguided by the literal meaning of spoken words. 
Furthermore, a fundamental limitation of SSL-based encoders \cite{hubert, Wavlm, wav2vec, wav2vec2.0} is their production of entangled representations. 
Within these learned features, affective prosody is inseparably mixed with phonetic content, rather than being explicitly disentangled. 
This conflation severely limits their robustness, particularly when faced with the challenge of affective-semantic conflict.
Our work directly addresses this representation entanglement and semantic bias by proposing a novel fusion framework.

\subsection{Neural Audio Tokenization}
The field of Text-To-Speech generation and Audio Editing has spurred the development of high-fidelity neural audio tokenizers. 
Discrete audio tokenizers, such as EnCodec \cite{Encodec}, XCodec \cite{Xcodec, Xcodec2} and VibeVoice \cite{VibeVoice}, built upon the VQ-VAE framework \cite{VQ-VAE}, excel at discretizing waveforms for high-quality signal reconstruction. More recently, VAE-based continuous tokenizers have also been proposed to better unify semantic and acoustic information for joint understanding and generation tasks \cite{Ming-UniAudio, DiTAR}, such as MingTok-Audio \cite{Ming-UniAudio}. 
A common characteristic of these generation-focused approaches is their ability to distill speech into a low-dimensional, acoustically-clean representation. Unlike the high-dimensional, semantically-entangled features from recognition-oriented encoders, this compact representation explicitly captures fine-grained prosody and speaker identity, which are crucial for high-quality synthesis.

While their primary application has been in audio generative tasks, their potential as a source of disentangled acoustic representation for discriminative tasks like SER has remained unexplored. 
We pioneer the repurposing of audio tokens as a dedicated pathway for modeling acoustic affective features, aiming to resolve emotional ambiguity in SER where traditional methods fail.

\section{Methods}

\begin{figure}[t]
  \includegraphics[width=\columnwidth]{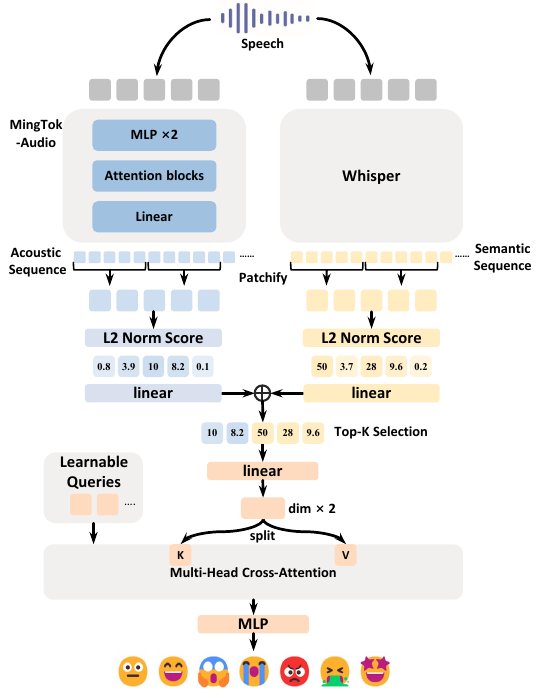}
  \caption{Overview of the proposed Fusion Acoustic-Semantic (FAS) framework. It disentangles speech into acoustic (MingTok-Audio) and semantic (Whisper) representations. The Fusion Module, guided by learnable queries and cross-attention to dynamically distill a unified feature for robust emotion recognition.}
  \label{fig:FAS_framework}
\end{figure}

\subsection{Overview of Fusion Acoustic-Semantic Framework}

The limitations of current SER models stem from their reliance on a single or entangled representation. Our core insight lies in effectively fusing two heterogeneous and temporally-varying feature streams: the semantic features $F_{\text{sem}} \in \mathbb{R}^{T_{\text{sem}} \times D_{\text{sem}}}$ and the acoustic features $F_{\text{aco}} \in \mathbb{R}^{T_{\text{aco}} \times D_{\text{aco}}}$, where $T$ and $D$ represent the sequence length and feature dimension, respectively.



As depicted in Figure~\ref{fig:FAS_framework}, we propose the \textbf{Fusion Acoustic-Semantic (FAS)}, a two-stage fusion framework that first intelligently distills salient tokens from feature streams and then bridges them using a cross-attention mechanism. 
The process is as follows:
\begin{enumerate}
    \item \textbf{Patchification:} To efficiently handle long sequence, we first apply a patchification step. Each sequence $F \in \mathbb{R}^{T \times D}$ is downsampled by a factor of $s=5$. This creates a shorter sequence of patches, $F' \in \mathbb{R}^{(T/s) \times D}$. 
    Subsequently, these patch sequences are projected into the unified hidden dimension, $d$:
    \begin{equation}
        f_{\text{aco}} = F'_{\text{aco}} W_{\text{aco}}; \quad f_{\text{sem}} = F'_{\text{sem}} W_{\text{sem}}
    \end{equation}
    where $f_{\text{aco}} \in \mathbb{R}^{T'_{\text{aco}} \times d}$ and $f_{\text{sem}} \in \mathbb{R}^{T'_{\text{sem}} \times d}$, with $T' = T/s$.

    \item \textbf{Token Distillation:} Recognizing that emotional cues are sparsely distributed, we introduce a non-uniform token selection strategy to identify and retain only the most informative "highlight" tokens from each sequence. This is achieved by:
        \begin{enumerate}
            \item \textbf{Saliency Scoring:} We compute a energy score $s_t$ for each token $f_t$ in a sequence. Inspired by findings that emotionally charged events often correlate with higher activation, we use the L2 Norm as a proxy for its score:
            \begin{equation}
                s_t = \| f_t \|_2
                \label{eq:saliency_score}
            \end{equation}
            This fast, non-parametric method effectively captures moments of high energy in the acoustic stream and text stream.
            
            \item \textbf{Top-K Selection:} We then select the $k$ tokens with the highest saliency scores, where $k_{\text{aco}}$ and $k_{\text{sem}}$ are sequence lengths chosen to reflect the different information densities of each pathway. This results in two condensed sequences:
            \begin{equation}
                \begin{aligned}
                    f'_{\text{aco}} &\in \mathbb{R}^{k_{\text{aco}} \times d} \\
                    f'_{\text{sem}} &\in \mathbb{R}^{k_{\text{sem}} \times d}
                \end{aligned}
            \end{equation}
        \end{enumerate}
    This distillation process drastically reduces sequence length while preserving the most emotionally relevant temporal information, which is a improvement over uniform compression techniques.

    \item \textbf{Learnable Queries:} The distilled sequences are concatenated to a context sequence $C \in \mathbb{R}^{(k_{\text{aco}} + k_{\text{sem}}) \times d}$. We then employ a fusion module inspired by Q-Former-like \cite{BLIP-2} architectures. A set of $n$ learnable queries, $Q_{\text{learn}} \in \mathbb{R}^{n \times d}$, actively interrogate this context to produce a learned-representation. The context $C$ is projected to generate Key ($K$) and Value ($V$) matrices, and a cross-attention mechanism computes the final fused tokens:
    \begin{align}
        F_{\text{fused}} = \text{Attn}(Q_{\text{learn}}, C W_K, C W_V)
    \label{eq:attention_fusion}
    \end{align}

\end{enumerate}

Finally, a simple Multi-Layer Perceptron (MLP) acts as the prediction head on top of the fused vector to yield the final emotion probabilities, $P(y|X)$, across the 7 emotion categories. 

\subsection{Conflict in Acoustic-Semantic Emotion (CASE) Benchmark}

\begin{figure}[t]
  \includegraphics[width=1.0\columnwidth]{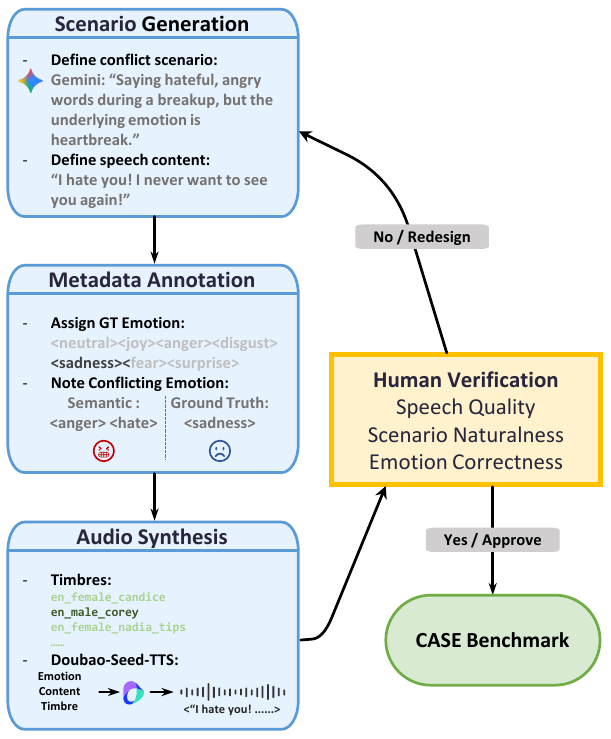}
  \caption{Pipeline of construction of CASE benchmark.}
  \label{fig:CASE}
\end{figure}

To rigorously evaluate methods robustness against affective-semantic conflict, we constructed the \textbf{Conflict in Acoustic-Semantic Emotion (CASE)} Benchmark, a specialized testbed designed to systematically probe SER model limitations in complex emotional scenarios.
The construction process in Figure~\ref{fig:CASE} was guided by the principles of logical coherence and high conflict density, ensuring that every sample is grounded in a plausible real-world scenarios.

The process began with scenario and text generation. Human experts, assisted by Gemini-2.5-pro \cite{Gemini,Gemini2.5}, crafted utterance texts that were laden with clear emotional sentiment. The emotion inherently implied by this verbal content was identified (e.g., \textit{angry} for the text "I'm going to give you one last chance") and served as the semantic anchor for creating the conflict.
Subsequently, in the metadata annotation stage, the core conflict was deliberately engineered. For each sample, we designated a ground-truth acoustic emotion with the strict constraint that it must conflict with the inherent semantic emotion of the text. 
To enhance diversity, a speaker timbre was also randomly selected from a pool of 21 multi-emotion voices.
This complete set of metadata—text, target emotion, and timbre—was then fed into the state-of-the-art TTS model Doubao-Seed-TTS 2.0 \cite{Seed-tts} for audio generation.

Finally, all synthesized samples underwent a rigorous manual verification by a panel of 12 human experts. The evaluation focused on whether the audio successfully projected the intended acoustic emotion with clarity, even when contradicted by the text. Samples were discarded if the acoustic prosody was perceived as weak, ambiguous, or overshadowed by the semantics. This stringent quality control process ensures that every sample in CASE presents a clear and challenging instance of affective conflict, culminating in a final benchmark of 378 high-quality samples.

\section{Experiments}

\subsection{Datasets}

\begin{table*}[t]
\centering

\small 
\setlength{\tabcolsep}{10pt} 

\begin{tabular}{l c c c c c c}
\toprule
\multicolumn{7}{c}{\textbf{Train \& In-Domain Test Sets}} \\
\midrule
\textbf{Dataset} & \textbf{\#Emo} & \textbf{Utts.} &\textbf{\#Hours} & \textbf{Train} & \textbf{Test} & \textbf{Lang} \\
\midrule

IEMOCAP \cite{IEMOCAP}  & 5     & 10,039    &   7.0   & \checkmark     & - & English \\
CMU-MOSEI \cite{CMU-MOSEI} & 7       & 5,239    & 9.3 & \checkmark         & - & English  \\
MER2024 \cite{MER2024} & 6     & 5,030    &   5.9   & \checkmark         & - &Multilingual       \\
MELD \cite{MELD}  & 7       & 13,847    & 12.2 & 11.2         & 1.0 & English  \\
RAVDESS \cite{RAVDESS} & 8       & 2452     &  2.8  & 2.3    & 0.5  &  English\\
ESD \cite{ESD}  & 5   & 17,500  &  29.0  &  23.7 & 5.3 & Multilingual\\
\midrule
\multicolumn{7}{c}{\textbf{Zero-Shot Test Sets}} \\
\midrule
\textbf{CASE} (Ours) & 7 & 378 &  0.32  & -     & \checkmark     &  Multilingual \\
Emo-Emilia \cite{C2SER}  & 7& 1400 &  3.29 & -   & \checkmark & Multilingual \\
EMOVO \cite{EMOVO}  & 7   & 588  & 0.51 & -    & \checkmark  &  Italian \\
EmoDB \cite{EmoDB} & 7   & 535   & 0.41  & - & \checkmark     &  German \\
\bottomrule
\end{tabular}
\caption{Overview of datasets used for training and evaluation. "\#Emo" denotes the number of emotion classes, "Utts." refers to the number of utterances, and "\#Hours" indicates the duration. The "Train" and "Test" columns specify the usage of each dataset in hours or with a “\checkmark” for full inclusion.}
\label{tab:datasets}
\end{table*}

Table~\ref{tab:datasets} provides an overview of the datasets used in our experiments, they are categorized into two groups: for training, in-domain testing, and for zero-shot evaluation.

To build a generalized model, we aggregated multiple open-source datasets into a large-scale, heterogeneous training corpus totaling over 66 hours. This includes MER2024 \cite{MER2024}, a multilingual video-based emotion recognition corpus; IEMOCAP \cite{IEMOCAP}, a dyadic conversational dataset of naturalistic emotional speech; CMU-MOSEI  \cite{CMU-MOSEI}, the largest sentiment analysis dataset with diverse topics and speakers; MELD \cite{MELD}, a TV dialogue dataset from Friends; RAVDESS \cite{RAVDESS}, a database of acted emotional speech and song by 24 professional actors; and ESD \cite{ESD}, a multilingual emotional speech dataset with 350 parallel utterances from 10 English and 10 Chinese native speakers.

Our evaluation is twofold. For in-domain testing, we use the official test splits of MELD, RAVDESS, and ESD to assess performance on distributions similar to training. More critically, our zero-shot evaluation (Table~\ref{tab:datasets}) measures language generalization on out-of-domain data. CASE serves as the primary testbed, designed to probe robustness against acoustic-semantic conflicts. We also include Emo-Emilia \cite{C2SER}, EMOVO \cite{EMOVO}, and EmoDB \cite{EmoDB}—covering Mandarin, Italian, and German—to evaluate cross-lingual and cross-corpus generalization, contrasting with the more spontaneous nature of our training data.
This rigorous zero-shot protocol is essential for verifying whether the model has learned transferable representations of emotion, rather than overfitting to the characteristics of the training sets.

\begin{table*}[ht]
\centering
\small
\setlength{\tabcolsep}{2pt} 

\begin{tabular}{ll | c | cc cc cc | cc cc cc cc}
\toprule
\multicolumn{2}{c|}{\multirow{2}{*}{\textbf{Modality}}} &
\multirow{3}{*}{\textbf{Methods}}  &\multicolumn{6}{c}{\textbf{In-Domain}} & \multicolumn{8}{|c}{\textbf{Zero-Shot}} \\
 & & &
\multicolumn{2}{c}{\multirow{1}{*}{\textbf{MELD}}} & 
\multicolumn{2}{c}{\multirow{1}{*}{\textbf{RAVDESS}}} & 
\multicolumn{2}{c|}{\multirow{1}{*}{\textbf{ESD}}} & 
\multicolumn{2}{c}{\multirow{1}{*}{\textbf{CASE}}} & 
\multicolumn{2}{c}{\multirow{1}{*}{\textbf{Emo-Emilia}}} & 
\multicolumn{2}{c}{\multirow{1}{*}{\textbf{EMOVO}}} & 
\multicolumn{2}{c}{\multirow{1}{*}{\textbf{EmoDB}}} \\

\textbf{Sem.} & \multicolumn{1}{r|}{\textbf{Aco.}} & &
\textbf{ACC} & \textbf{F1} & 
\textbf{ACC} & \textbf{F1} & 
\textbf{ACC} & \textbf{F1} &
\textbf{ACC} & \textbf{F1} &
\textbf{ACC} & \textbf{F1} &
\textbf{ACC} & \textbf{F1} &
\textbf{ACC} & \textbf{F1} \\
\midrule
\multicolumn{11}{l}{\textit{SSL Models}} \\
\multicolumn{1}{c}{\checkmark} & \multicolumn{1}{c|}{\checkmark} & HuBERT & 45.99&38.07&69.96&68.79&80.10&79.13&32.90&32.24&34.36&29.95&29.05&19.10&52.99&53.57\\
\multicolumn{1}{c}{\checkmark} & \multicolumn{1}{c|}{\checkmark} & WavLM  & 44.64&34.09&61.90&60.69&77.43&76.63&34.20&33.92&35.64&29.33&30.74&21.40&64.18&56.74\\
\multicolumn{1}{c}{\checkmark} & \multicolumn{1}{c|}{\checkmark} & wav2vec 2.0  & 37.44&28.36&23.59&16.44&21.81&19.57&25.59&22.83&19.64&16.02&21.11&15.28&25.37&22.09\\
\midrule
\multicolumn{11}{l}{\textit{Semantic Models}} \\
\checkmark & \multicolumn{1}{c|}{} & Whisper  & 49.59&43.62&62.30&60.61&84.53&\underline{83.92}& \underline{47.26} & \underline{44.97} &50.50&42.29&\textbf{43.07}&\textbf{34.38}&68.84&63.46
\\
\checkmark & \multicolumn{1}{c|}{} & CLAP & 43.74&34.96&47.38&44.83&59.97&59.40&34.46&31.93&24.93&18.83&34.63&26.57&43.28&41.45\\
\midrule
\multicolumn{11}{l}{\textit{Neural Audio Tokenizers}} \\
& \multicolumn{1}{c|}{\checkmark} & EnCodec  & 34.38&29.41&22.78&18.44&28.73&25.18&24.54&18.81&15.64&11.47&24.66&18.33&29.10&23.87
 \\
&  \multicolumn{1}{c|}{\checkmark} & Vibevoice & 39.06&30.43&37.90&31.48&37.61&36.40&30.03&25.90&19.36&15.35&22.47&17.95&28.73&20.92
 \\
&  \multicolumn{1}{c|}{\checkmark} & MingTok-Audio & 41.94&29.76&36.49&26.12&29.93&28.80&29.24&25.61&21.07&16.34&21.45&16.34&37.31&34.02
\\
\midrule
\multicolumn{11}{l}{\textit{Other Open-Sourced SER Methods}} \\
\multicolumn{2}{c|}{} & Emotion2Vec & 45.04 & 45.49 & 70.06 & 68.84 & 51.39 & 50.87 & 31.48 & 28.42 & 52.79 & 50.44 & 33.53 & 29.01 & 71.21 & \underline{76.07}\\
\multicolumn{2}{c|}{} & Vesper \(^{\dagger}\) & 25.00  & \underline{45.70} & - & - & - & - & - & - &- &- &- &- & - &-\\
\midrule
\multicolumn{11}{l}{\textit{Audio Language Models}} \\
\checkmark & \checkmark  & $\text{C}^{2} \text{SER}$ \(^{\dagger}\) & 51.39 & 27.45 & -&- & \textbf{93.86 }& 68.19 &- & -& 68.29 & 61.28 & 37.59& 27.33 &-&-\\
\checkmark & \multicolumn{1}{c|}{} & Qwen2-Audio & 35.29 & 29.91 &\textbf{85.74}&\textbf{86.59}& 36.99 & 23.35 &32.53&27.08 &\underline{69.64} & \textbf{68.81}&26.87& 20.29& \underline{74.21} & 70.29 \\
\checkmark & \multicolumn{1}{c|}{} & Qwen2.5-Omni &\textbf{54.06} & 36.05 & 75.35 & 74.98 &51.60&35.70 & 34.66 & 30.21 & \textbf{70.64 }& \underline{68.03} & 27.89 & 20.03 & \textbf{87.85 }&\textbf{85.97} \\
\midrule

\checkmark & \multicolumn{1}{c|}{\checkmark} & \textbf{FAS (Ours)} & \underline{51.89} &\textbf{48.42}& \underline{76.61} & \underline{76.19} & \underline{87.27} & \textbf{86.72} & \textbf{59.38} &\textbf{55.08}&51.14&42.92& \underline{40.03 }& \underline{33.39} & 68.10 &65.07 \\
\bottomrule
\end{tabular}
\caption{In-domain and Zero-shot generalization performance across all datasets. The table is categorized by model paradigm. The first two columns indicate whether the baseline model primarily captures Semantic (\textbf{Sem.}) or Acoustic (\textbf{Aco.}) information. For each benchmark, we report both Accuracy (\textbf{ACC}) and F1 Score (\textbf{F1}). \textbf{AVG} represents the average ACC and F1 across these three in-domain sets. "\(^{\dagger}\)" denotes their results are from the official Versper \cite{Vesper} and $\text{C}^{2} \text{SER} \text{(Explicit CoT)}$  \cite{C2SER} paper. The best results are \textbf{bolded} and the second-best results are \underline{underlined}.}
\label{tab:comparison}
\end{table*}

\subsection{Implementation Details}
\label{sec:implementation_details}

\begin{table}[h]
\centering
\small 

\begin{tabular}{l c c}
\toprule
\textbf{Hyperparameters} & \textbf{FAS} & \textbf{Concat\&Gated} \\
\midrule
Hidden Dimension ($d$) & 512 & 512 \\
Query Length ($N_q$) & 2 & -\\
Dropout Rate & 0.4 & 0.4\\
\midrule
Optimizer & \multicolumn{2}{c}{AdamW} \\
Learning Rate & \multicolumn{2}{c}{$2 \times 10^{-4}$} \\
LR Schedule & \multicolumn{2}{c}{Cosine} \\
Weight Decay & \multicolumn{2}{c}{$1 \times 10^{-4}$} \\
Global Batch Size  & \multicolumn{2}{c}{2048} \\
Loss & \multicolumn{2}{c}{Cross-Entropy} \\
Epochs & \multicolumn{2}{c}{100} \\
Warmup Ratio & \multicolumn{2}{c}{0.05} \\
Sample Rate & \multicolumn{2}{c}{16000} \\
\bottomrule
\end{tabular}
\caption{Hyperparameter settings for the experiments. The `Concat \& Gated` column specifies the shared parameters for the Concatenation and Gated Fusion baselines, which are used for the ablation studies in Section~\ref{sec:Efficacy_of_FAS_framework}.}
\label{tab:hyperparameters}
\end{table}

Our experiments were conducted on $8$ $\times$ NVIDIA A6000 GPUs. 
A fixed random seed of $42$ was used for all experiments to ensure reproducibility.
For the Semantic Pathway, we use the encoder from the pre-trained Whisper-large \cite{whisper} model to extract a 1280-dimensional feature.
For the Acoustic Pathway, we employ the MingTok-Audio \cite{Ming-UniAudio} tokenizer to extract a 64-dimensional feature.
Then our proposed FAS is trained from scratch on these pre-computed features. This approach accelerates experimentation by decoupling the heavy feature extraction from the training of the lightweight fusion module.
As shown in Table~\ref{tab:hyperparameters}, the fusion module is configured with a unified hidden dimension of $d=512$. The entire model is trained end-to-end using the AdamW optimizer \cite{AdamW} with an initial learning rate of $2 \times 10^{-4}$ and a weight decay of $1 \times 10^{-4}$. 
We use a global batch size of $2048$ and train for $100$ epochs, with Cross-Entropy Loss as the optimization objective. 

\subsection{Evaluation Metrics and Comparison baselines}
\label{sec:eval and baselines}

We employ two standard metrics for the evaluation: Accuracy (ACC) and Unweighted Average F1 score. Accuracy provides a global measure of correctness, while F1 is crucial for assessing performance on imbalanced datasets.

Our comparative evaluation includes a rigorous selection of strong baselines across diverse representational paradigms. 
Our baselines span multiple paradigms: (1) self-supervised speech models (HuBERT~\cite{hubert}, WavLM~\cite{Wavlm}, wav2vec 2.0~\cite{wav2vec2.0}); (2) large-scale speech-text pre-trained encoders (Whisper~\cite{whisper}, CLAP~\cite{CLAP}); (3) neural audio tokenizers for TTS (EnCodec~\cite{Encodec}, VibeVoice~\cite{VibeVoice}, MingTok-Audio~\cite{Ming-UniAudio}); (4) current SOTA SER methods (Emotion2Vec~\cite{emotion2vec}, Vesper~\cite{Vesper});
and (5) Audio Language Models (ALMs) such as $\text{C}^2\text{SER}$ 7B \cite{C2SER}, Qwen2-Audio-Instruct 7B \cite{Qwen2-VL} and Qwen2.5-Omni 7B \cite{Qwen2_5_omni}, evaluated via their official pipeline with standardized emotion prompts.

To ensure a fair and rigorous comparison across these baselines, for all non-ALM models, we freeze the pre-trained encoder, extract mean-pooled utterance embeddings, and train a lightweight two-layer classifier. 
For task-specific models trained on different emotion taxonomies, their outputs are projected onto a unified label space consistent with our target benchmarks for comparison.

\subsection{Main Results}

We evaluate FAS on both in-domain (MELD, RAVDESS, ESD) and zero-shot (CASE, Emo-Emilia, EMOVO, EmoDB) settings. As shown in Table~\ref{tab:comparison}, FAS achieves state-of-the-art results across the board: it obtains an average in-domain ACC of \textbf{71.92\%}, outperforming SSL models, semantic encoders, audio tokenizers, and even large audio-language models (ALMs) like Qwen2-Audio and Qwen2.5-Omni. Notably, while Qwen2.5-Omni excels on high-resource datasets (e.g., 87.85\% ACC on EmoDB), it underperforms on challenging zero-shot benchmarks such as CASE (34.66\%) and EMOVO (27.89\%). In contrast, FAS delivers robust performance—reaching \textbf{59.38\% SOTA ACC on CASE} and \textbf{54.66\% average ACC} across all zero-shot tasks—demonstrating its ability to generalize under distribution shift. 
This performance gap reveals a fundamental trade-off in current ALMs: their architecture is optimized for alignment with the LLM backbone, which emphasizes textual semantics while drop the affective nuances carried by acoustic prosody.
By explicitly modeling interactions between prosody and semantics, FAS bridges this gap, enabling reliable emotion recognition in both familiar and unseen scenarios.
More experimental results including loss curves, confusion matrices, and visualizations of features map are provided in the \textit{Appendix Section~\ref{sec:additional_analyses}}.

\subsection{Ablation Study}
\label{sec:ablation_study}

To validate the superiority within our FAS framework, we conduct a series of comprehensive ablation studies to answer three central questions:

(1) How is the proposed FAS compared to other strategies?
(2) Is the FAS framework a generalizable "plug-and-play" solution, or is its success tied to a specific model pair?
(3) How does the internal configuration of the FAS influence its ability?

\subsubsection{Efficacy of FAS framework}
\label{sec:Efficacy_of_FAS_framework}

\begin{table}[h]
\centering
\small
\setlength{\tabcolsep}{2pt}
\begin{tabular}{ l| c | cc  cc cc }
\toprule
\multirow{2}{*}{\textbf{Methods}} 
& \multirow{2}{*}{\textbf{Param}} 
& \multicolumn{2}{c}{\textbf{CASE}} 
& \multicolumn{2}{c}{\textbf{MELD}} 
& \multicolumn{2}{c}{\textbf{RAVDESS}} \\
& 
& \textbf{ACC} & \textbf{F1} 
& \textbf{ACC} & \textbf{F1} 
& \textbf{ACC} & \textbf{F1} \\
\midrule
Concat & 1.22M & 53.65&50.77&52.88&48.50&75.40&75.22\\
Gated  & 1.65M & 53.12&50.84&\textbf{52.70}&47.92&73.79&73.59 \\
\midrule
w/o Top-K & 3.45M & 55.47&51.57 & 52.70&\textbf{48.60}&73.79&73.32\\
w/o $Q_{learn}$ & 0.82M & 55.99&52.48 &48.65&44.71&67.34&66.95\\
FAS    & 3.45M &\textbf{59.38} &\textbf{55.08} &51.89&48.42 &\textbf{76.61}&\textbf{76.19}\\
\bottomrule
\end{tabular}
\caption{Ablation study of fusion methods on CASE (zero-shot), MELD, and RAVDESS, all built upon Whisper and MingTok-Audio encoders. "w/o Top-K" removes token selection (uses random token); "w/o $Q_{\text{learn}}$" removes learnable queries. Best results are \textbf{bolded}.}
\label{tab:ablation_on_fusion}
\end{table}

\begin{figure}[t]
  \includegraphics[width=1.0\columnwidth]{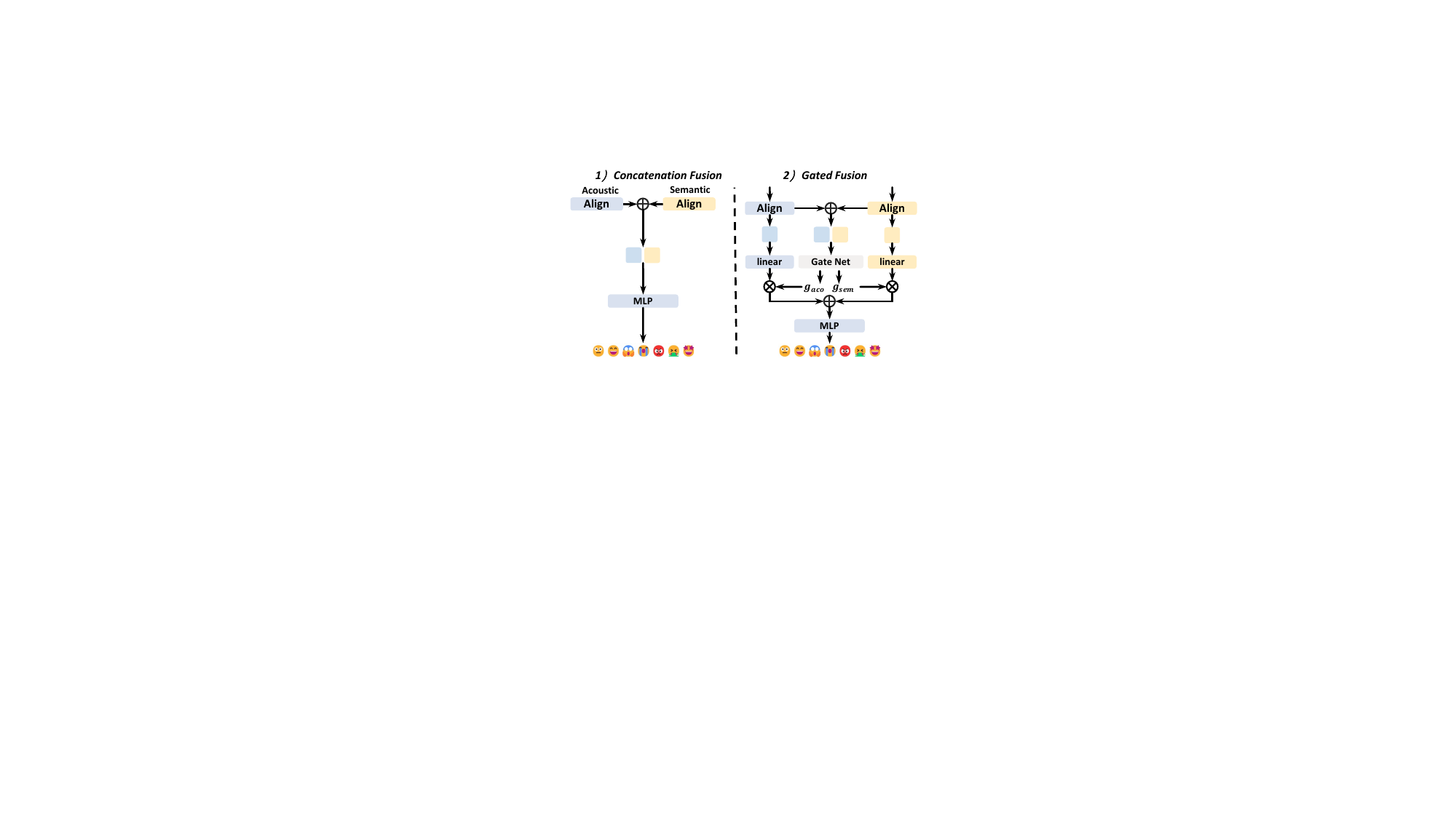}
  \caption{Illustration of different fusion strategies, including Concatenation Fusion and Gated Fusion.}
  \label{fig:ablation}
\end{figure}



To validate the core design of the FAS framework, we compare against both classical fusion baselines and key architectural ablations. As shown in Table~\ref{tab:ablation_on_fusion}, naive strategies like concatenation or gating offer limited gains, confirming that passive combination is insufficient for cross-type feature integration. 
Critically, removing either component of FAS leads to significant performance drops: 
(1) \textit{w/o Top-K}—which random select tokens without energy scores—underperforms FAS by up to 3.91\% ACC on CASE; 
(2) \textit{w/o $Q_{\text{learn}}$}—which replaces learnable queries with original inputs, suffers a severe drop on RAVDESS (-9.27\% ACC). 
FAS achieves consistent gains over strong baselines with only a negligible increase in parameters, demonstrating that its superiority stems from the synergistic design of token distillation and learnable queries.

\subsubsection{Framework Generalizability}

\begin{table*}[htbp]
\centering
\small
\renewcommand{\arraystretch}{1.15}
\setlength{\tabcolsep}{3pt}
\begin{tabular}{cc  | cc cc cc | cc cc cc cc | cc}
\toprule
\multirow{2}{*}{\textbf{$k_{aco}$}} & \multirow{2}{*}{\textbf{$k_{sem}$}} 
& \multicolumn{2}{c}{\textbf{MELD}} 
& \multicolumn{2}{c}{\textbf{RAVDESS}} 
& \multicolumn{2}{c}{\textbf{ESD}} 
& \multicolumn{2}{|c}{\textbf{CASE}} 
& \multicolumn{2}{c}{\textbf{Emo-Emilia}} 
& \multicolumn{2}{c}{\textbf{EMOVO}} 
& \multicolumn{2}{c}{\textbf{EmoDB}} 
& \multicolumn{2}{|c}{\textbf{AVG}} \\
& & \textbf{ACC} & \textbf{F1} 
& \textbf{ACC} & \textbf{F1} 
& \textbf{ACC} & \textbf{F1} 
& \textbf{ACC} & \textbf{F1} 
& \textbf{ACC} & \textbf{F1} 
& \textbf{ACC} & \textbf{F1} 
& \textbf{ACC} & \textbf{F1} 
& \textbf{ACC} & \textbf{F1} \\
\midrule
8 & 8  & 51.62&47.99&77.82&77.51&86.84&86.25&58.59&54.58&50.64&42.66&38.68&31.57&66.23&62.47&61.49 & 57.58\\
16 & 16  & 51.89&48.35&77.02&76.59&87.20&86.72&56.77&52.70&51.29&42.84&39.36&32.11&66.79&64.27&61.47&57.65\\
\midrule
16 & 8  & 51.35&47.73&78.23&77.75&87.16&86.66&56.25&52.53&51.21&43.05&38.85&31.14&67.16&64.22&61.46&57.58\\
8 & 16  & 51.89&48.42&76.61&76.19&87.27&86.72& \textbf{59.38} &\textbf{55.08}&51.14&42.92&40.03&33.39& \textbf{68.10} &\textbf{65.07}&\textbf{62.06}&\textbf{58.26}\\
\midrule
32 & 16  & 50.90&47.44&78.83&78.47&86.70&86.40&55.21&51.47&\textbf{52.36}&\textbf{43.80}&\textbf{41.89}& \textbf{35.30} & 64.74 & 61.88 & 61.52 & 57.82\\
16 & 32  & \textbf{52.70}&\textbf{49.11}&\textbf{79.03}&\textbf{78.77}&\textbf{87.66}&\textbf{87.20}&56.25&52.37&52.29&43.76&39.70&32.04&64.37&61.98&61.71&57.89\\

\bottomrule
\end{tabular}
\caption{Ablation on FAS hyper-parameters:  acoustic sequence length ($k_{\text{aco}}$), semantic sequence length ($k_{\text{sem}}$). Hidden dimension ($d=512$) and the length of learnable queries ($N_q=2$) are fixed across all benchmarks. Best results per dataset are marked \textbf{bolded}.}
\label{tab:hyperparam_ablation}
\end{table*}



To demonstrate the generalization of our FAS framework, we evaluate its ``plug-and-play'' capability with diverse acoustic and semantic backbones. 
For the \textbf{acoustic pathway}, we substitute MingTok-Audio with VibeVoice and XCodec2. For the \textbf{semantic pathway}, we compare Whisper and CLAP—two contrastively trained models with distinct training objectives.
As shown in Table~\ref{tab:encoder_generalization}, FAS consistently enables strong cross-modal fusion across all combinations. FAS outperforms single-pathway baselines (marked with ``–''), confirming that gains stem from fusion rather than individual encoders. 
FAS w/ (Whisper+XCodec2) achieves the best MELD performance (52.34\% ACC), while FAS w/ (Whisper+VibeVoice) yields the highest RAVDESS score (80.04\% ACC)—both surpassing our default MingTok+Whisper pairing on their respective datasets. On the zero-shot CASE benchmark, MingTok+Whisper remains optimal (59.38\% ACC), suggesting that tokenized continuous acoustic tokens better support cross-lingual transfer when paired with a strong semantic encoder.
These results confirm that FAS effectively bridges heterogeneous features, and its performance scales with encoder quality rather than being tied to a fixed encoder pair.

\begin{table}[h]
\centering

\small
\renewcommand{\arraystretch}{1.2}
\setlength{\tabcolsep}{2pt} 
\begin{tabular}{c c | c c c}
\toprule
\multirow{2}{*}{\textbf{Acoustic}}  & \multirow{2}{*}{\textbf{Semantic}}  & \textbf{CASE} & \textbf{MELD} & \textbf{RAVDESS} \\
& & \tiny{(ACC / F1)} & \tiny{(ACC / F1)} & \tiny{(ACC / F1)} \\
\midrule
- & Whisper & 47.26/44.97&49.59/43.62&62.30/60.61 \\
Vibevoice & Whisper  &58.07/53.35 & 51.53/48.06 & \textbf{80.04}/\textbf{79.71}\\
XCodec2 & Whisper &58.33/54.46& \textbf{52.34}/\textbf{48.87}&79.03/78.76 \\
\midrule
- & CLAP& 34.46/31.93 &43.74/34.96&47.38/44.83 \\
MingTok & CLAP   &33.85/31.03 & 40.83/36.06 & 62.50/62.04\\
Vibevoice & CLAP& 36.72/34.19& 35.43/34.07&63.31/62.72\\
XCodec2 & CLAP & 32.55/30.63 & 43.26/38.14&59.07/58.81\\
\midrule
\textbf{MingTok} & \textbf{Whisper} & \textbf{59.38} /\textbf{55.08} &51.89/48.42 &76.61/76.19 \\
\bottomrule
\end{tabular}
\caption{Generalization of the FAS framework across diverse acoustic and semantic encoders on CASE (zero-shot), MELD, and RAVDESS. Results demonstrate that FAS consistently enables effective cross-type fusion regardless of encoder architecture. }
\label{tab:encoder_generalization}
\end{table}

\subsubsection{Ablation on Hyper-parameters of FAS}

\begin{figure}[t]
  \includegraphics[width=1.0\columnwidth]{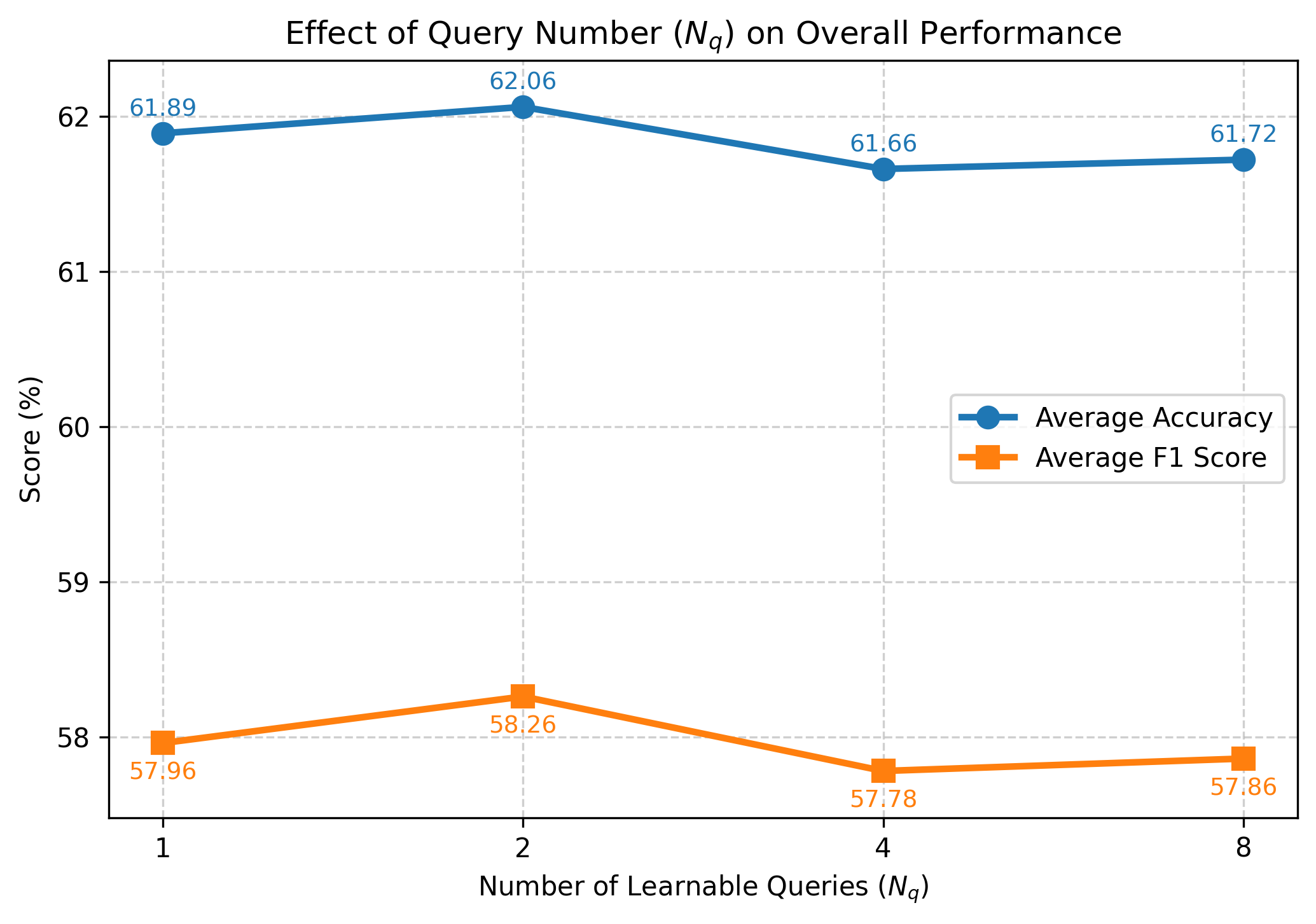}
  \caption{Ablation on the number of learnable queries ($N_q$). Average Accuracy and F1 scores are computed across all datasets.}
  \label{fig:query_ablation}
\end{figure}

Table~\ref{tab:hyperparam_ablation} presents a systematic ablation on the number of retained acoustic ($k_{\text{aco}}$) and semantic ($k_{\text{sem}}$) tokens. 
Specifically, increasing $k_{\text{sem}}$ (e.g., from 8 to 16) consistently improves performance on zero-shot benchmarks—most notably on CASE (+0.79\% ACC) and EmoDB (+1.35\% ACC)—suggesting that richer semantic context enhances cross-lingual and cross-corpus transfer. In contrast, enlarging $k_{\text{aco}}$ provides marginal or even negative gains on in-domain datasets such as MELD and RAVDESS, indicating diminishing returns from redundant acoustic frames.
The best overall average performance (62.06\% ACC) is achieved with \textbf{$k_{\text{aco}} = 8$, $k_{\text{sem}} = 16$}, revealing an asymmetric design principle: \textit{preserving more semantic tokens is more beneficial than retaining additional acoustic ones} for both in-domain and zero-shot settings. 

To further investigate the role of learnable queries ($N_q$), we conducted an ablation study varying $N_q$ from 1 to 8.
As shown in Figure~\ref{fig:query_ablation}, we find that performance saturates at $N_q = 2$, with the best average ACC (62.06\%) and F1 (58.26\%). Increasing $N_q$ to 4 or 8 yields no gain—often slight degradation—while $N_q = 1$ already achieves competitive results (61.89\% ACC). This confirms that SER, as a low-complexity utterance-level classification task, requires only minimal query capacity; additional queries introduce redundancy without improving generalization. 

\section{Conclusion}
In this work, we address an overlooked challenge in SER-Acoustic-Semantic Conflict, where acoustic prosody conveys an emotion that contradicts the literal meaning. 
We demonstrate that current SER methods—ranging from ASR-based to SSL models—are brittle in such scenarios due to semantic bias or entangled representations.
To tackle this, we propose the \textbf{FAS} framework, which explicitly disentangles and bridges acoustic and semantic pathways using a query–based fusion module. Along with the \textbf{CASE}, the first benchmark specifically designed to evaluate model robustness under emotional conflicts. Extensive experiments show FAS framework outperforms SOTA baselines across both in-domain and zero-shot settings. 
While FAS demonstrates strong performance as a lightweight SER method, its potential as an integrated component within end-to-end ALMs remains unexplored, which could be investigated in the future work.



\newpage

\section*{Limitations}
\label{sec:limitations}

While our work introduces a novel perspective on robust speech emotion recognition under acoustic-semantic conflict, several limitations delineate the current scope of our investigation and suggest promising avenues for future research.
First, although the CASE benchmark incorporates multiple languages—including English, Mandarin, and representative Chinese dialects—its linguistic coverage remains limited. The phenomena of acoustic-semantic conflict may manifest differently across a broader range of language families, tonal systems, or cultural contexts. 
Second, CASE is designed primarily as a diagnostic evaluation suite, not a large-scale training resource. With fewer than 400 high-quality, human-verified conflict samples, it provides a controlled testbed for probing model robustness but is insufficient in scale to serve as standalone training data.


Finally, our current formulation centers on the binary tension between acoustic and semantic signals, which captures a prevalent and impactful class of conflicts (e.g., sarcasm, polite masking). However, real-world emotional expression can involve additional contextual cues—such as speaker identity or conversational history—that are not explicitly modeled in FAS. Incorporating these richer signals could enable even more nuanced conflict resolution in future systems.

\section*{Ethical Considerations}
\label{sec:ethical_considerations}

We have taken several measures to ensure the ethical integrity of this research.
All source datasets used for model training (IEMOCAP, MELD, CMU-MOSEI, etc.) are publicly available academic corpora that have been widely adopted in the affective computing community. Our newly introduced CASE benchmark is entirely synthetic, generated from scripted text prompts using a commercial-grade TTS engine. Consequently, it contains no personally identifiable information (PII) or recordings of real individuals, thereby mitigating privacy concerns associated with collecting sensitive emotional data.

The human verification process for CASE involved 12 expert annotators—recruited from our institution’s pool of linguistics co-workers—who were provided with clear annotation guidelines and compensated at a standard academic rate, which is fair and adequate for their demographic and task complexity.
Their task was limited to evaluating the perceptual quality and emotional clarity of the synthetic audio, not to disclose any personal information.

We acknowledge the potential for misuse of robust SER technology. A system capable of accurately inferring true emotions despite verbal content could be deployed in surveillance, manipulative advertising, or high-stakes interrogation settings without the subject's consent. 
To mitigate these risks, we emphasize that our work is intended solely for research purposes to improve the fundamental understanding of emotion expression. We advocate for the development and enforcement of strict ethical guidelines and regulatory frameworks governing the deployment of such technologies in real-world applications, ensuring user consent, transparency, and the right to opt-out.



\bibliography{custom}

\appendix

\section{Appendix}
\label{sec:appendix}

\subsection{Case Study}
\label{sec:case_study}

\begin{figure*}[ht]
    \centering
    \includegraphics[width=\linewidth]{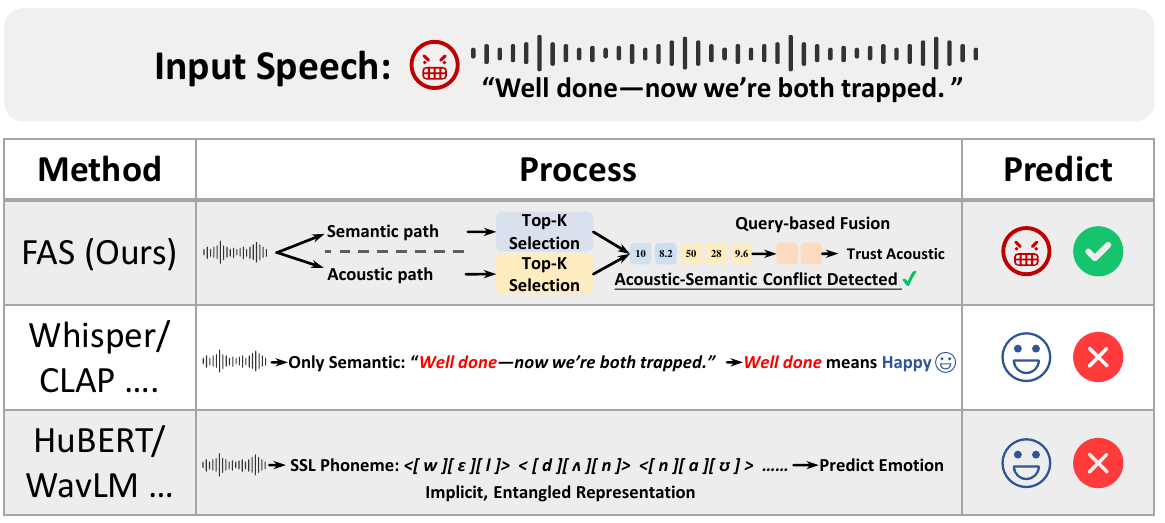}
    \caption{A comparison of SER methods on a conflicting utterance. FAS (Ours) explicitly disentangles acoustic and semantic pathways, detects their conflict via query-based fusion, and prioritizes the acoustic signal to correctly predict emotion.}
    \label{fig:emotion_comparison}
\end{figure*}


To illustrate the challenge of acoustic-semantic conflict in emotion recognition, consider the example shown in Figure~\ref{fig:emotion_comparison}: ``Well done—now we’re both trapped.'' While the lexical content ("well done") typically signals positive sentiment, the speaker’s angry tone reveals a negative emotional state. This discrepancy poses a significant challenge for existing models.

Previous approaches such as Whisper, which primarily rely on semantic understanding derived from text transcripts, are prone to misclassification due to their strong language priors. They interpret "well done" as inherently positive, leading to an incorrect prediction of happiness (see Figure~\ref{fig:emotion_comparison}).
Similarly, SSL models like HuBERT and WavLM, though capable of extracting fine-grained acoustic features, operate on entangled representations where phonetic and prosodic information are not cleanly separated. 
As a result, even when they capture the angry prosody, the model lacks explicit mechanisms to resolve the conflict cues, defaulting to ambiguous predictions.

In contrast, our proposed \textbf{FAS framework} addresses this issue by explicitly modeling two parallel pathways. Through Top-K selection and query-based fusion, FAS detects discrepancies and applies a confidence-aware decision rule.

\subsection{Additional Analyses}
\label{sec:additional_analyses}

\begin{figure*}[htbp]
\centering
\includegraphics[width=0.9\textwidth]{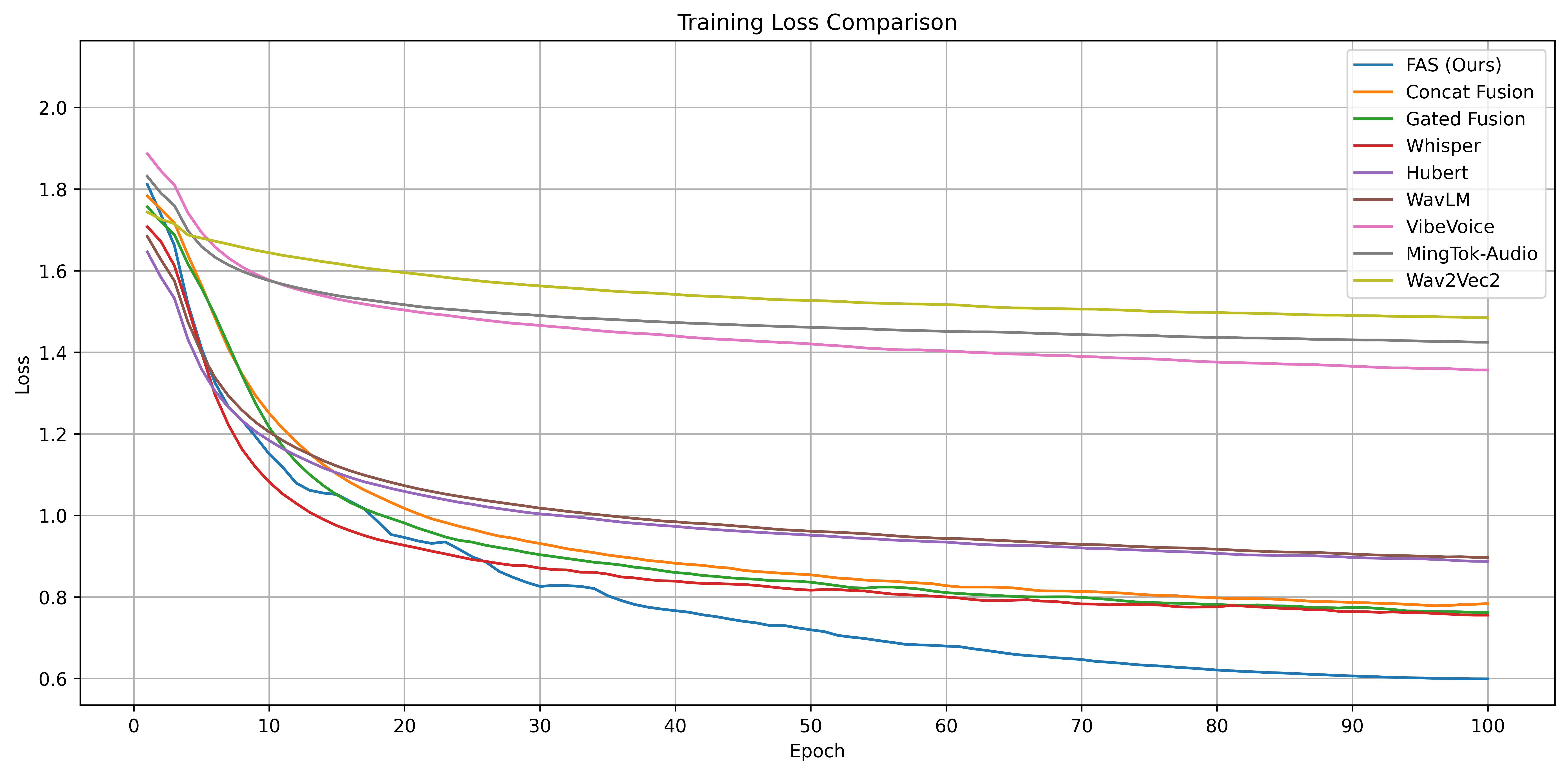}
\caption{Training loss curves for different SER strategies. FAS demonstrates slower convergence but lower final loss.}
\label{fig:loss_curves}
\end{figure*}

\subsubsection{Loss Curve}
\label{sec:loss_curves}

To better understand the learning behavior of our proposed \textbf{Fusion Acoustic-Semantic (FAS)} framework compared to alternative strategies, we plot the training loss curves. 

Figure~\ref{fig:loss_curves} shows the training loss trajectories of multiple methods, including our proposed \textbf{FAS}, its ablation variants (w/ Concatenation Fusion and Gated Fusion), and several strong baselines.
The results reveal that while FAS exhibits slower initial convergence, it eventually achieves a significantly lower final loss plateau compared to other methods. In contrast, both Concatenation and Gated Fusion converge more rapidly in the early stages but stabilize at higher loss values, indicating potentially poorer generalization.
This delayed convergence may be attributed to the complexity of learning alignment through the $Q_{learn}$ fusion module, which requires more epochs to stabilize. However, once optimized, the learned representations demonstrate superior discriminative power.

\subsubsection{Space Visualization}
\label{sec:visualization}

To qualitatively assess the effectiveness of our FAS framework in learning discriminative emotion representations, we visualize the utterance-level embeddings using Uniform Manifold Approximation and Projection (UMAP)~\cite{UMAP}. 
Features are extracted from the final hidden layer of each model and projected onto a 2D space for comparison on the CASE benchmark.

\begin{figure*}[htbp]
\centering
\includegraphics[width=\textwidth]{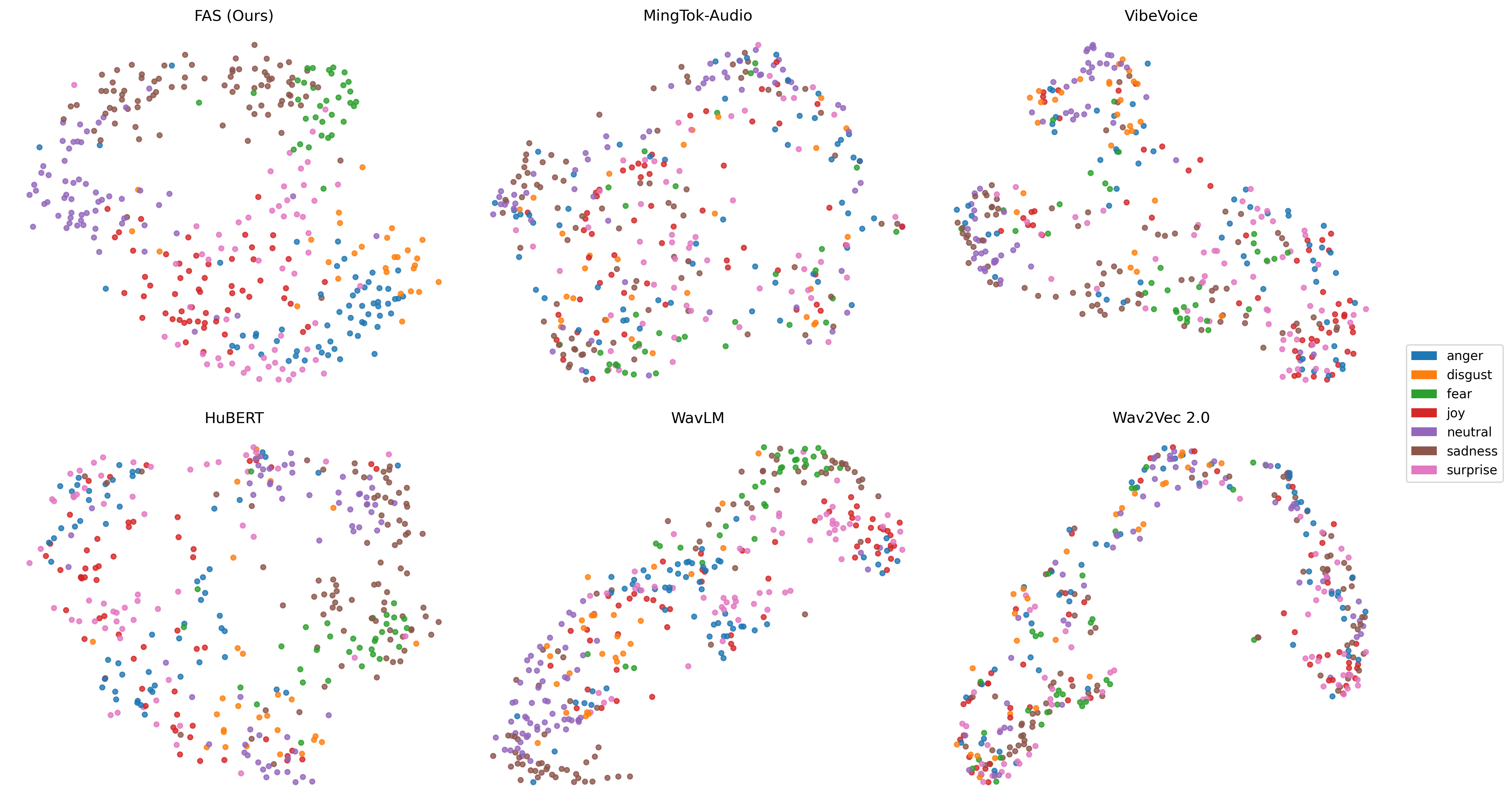}
\caption{UMAP visualization of deep representations on CASE. Colors denote ground-truth acoustic emotion labels. FAS (top-left) achieves the most distinct emotion clusters, whereas HuBERT (bottom-left) and Wav2Vec 2.0 (bottom-right) show significant overlap.}
\label{fig:umap_CASE}
\end{figure*}

As shown in Figure~\ref{fig:umap_CASE}, baseline models such as HuBERT \cite{hubert} and Wav2Vec 2.0 \cite{wav2vec2.0} exhibit highly mixed and indistinct clusters, indicating limited ability to disentangle emotional content from other factors.
In contrast, our FAS framework yields clearly separated and well-structured clusters that strongly correlate with the ground-truth acoustic emotion labels.

\begin{figure*}[htbp]
\centering
\includegraphics[width=\textwidth]{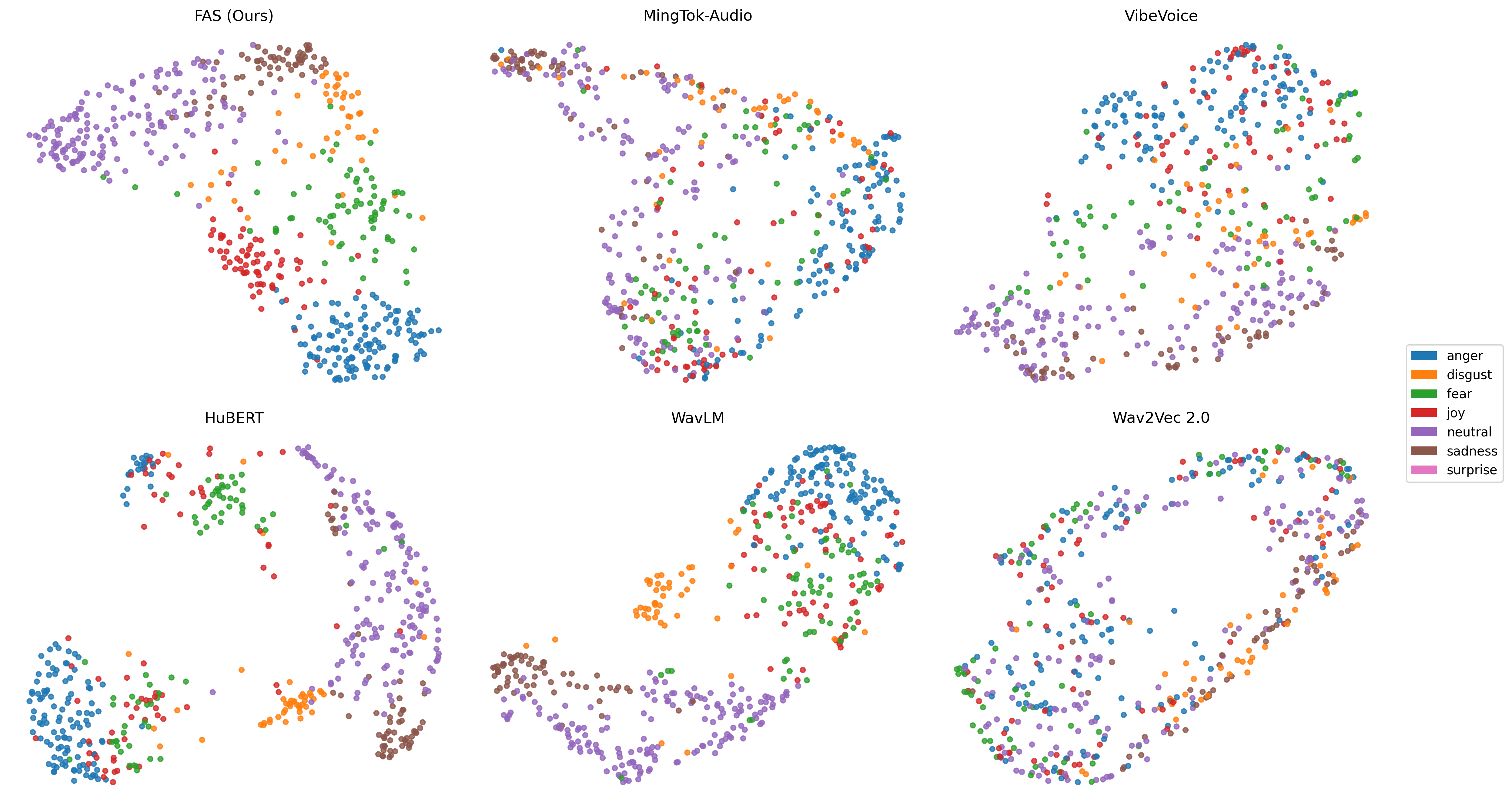}
\caption{UMAP visualization of deep representations on EmoDB, a German emotional speech dataset. Colors denote ground-truth acoustic emotion labels. Despite the challenges of cross-lingual generalization and limited data size, our FAS framework still produces relatively structured clusters compared to baseline models, demonstrating its robustness in diverse conditions.}
\label{fig:umap_EmoDB}
\end{figure*}

We further evaluate our framework on EmoDB \cite{EmoDB}, a German emotional speech corpus with limited data size and lower recording quality. As shown in Figure~\ref{fig:umap_EmoDB}, while the overall clustering structure is less pronounced due to these challenges, our FAS framework still maintains relatively coherent emotion clusters, outperforming most baselines.
This demonstrates the robustness of our approach across different languages and data conditions, reinforcing its practical applicability.


\subsubsection{Confusion Matrix Analysis}

\begin{figure*}[htbp]
    \centering
    \begin{subfigure}[b]{0.30\textwidth}
        \includegraphics[width=\linewidth]{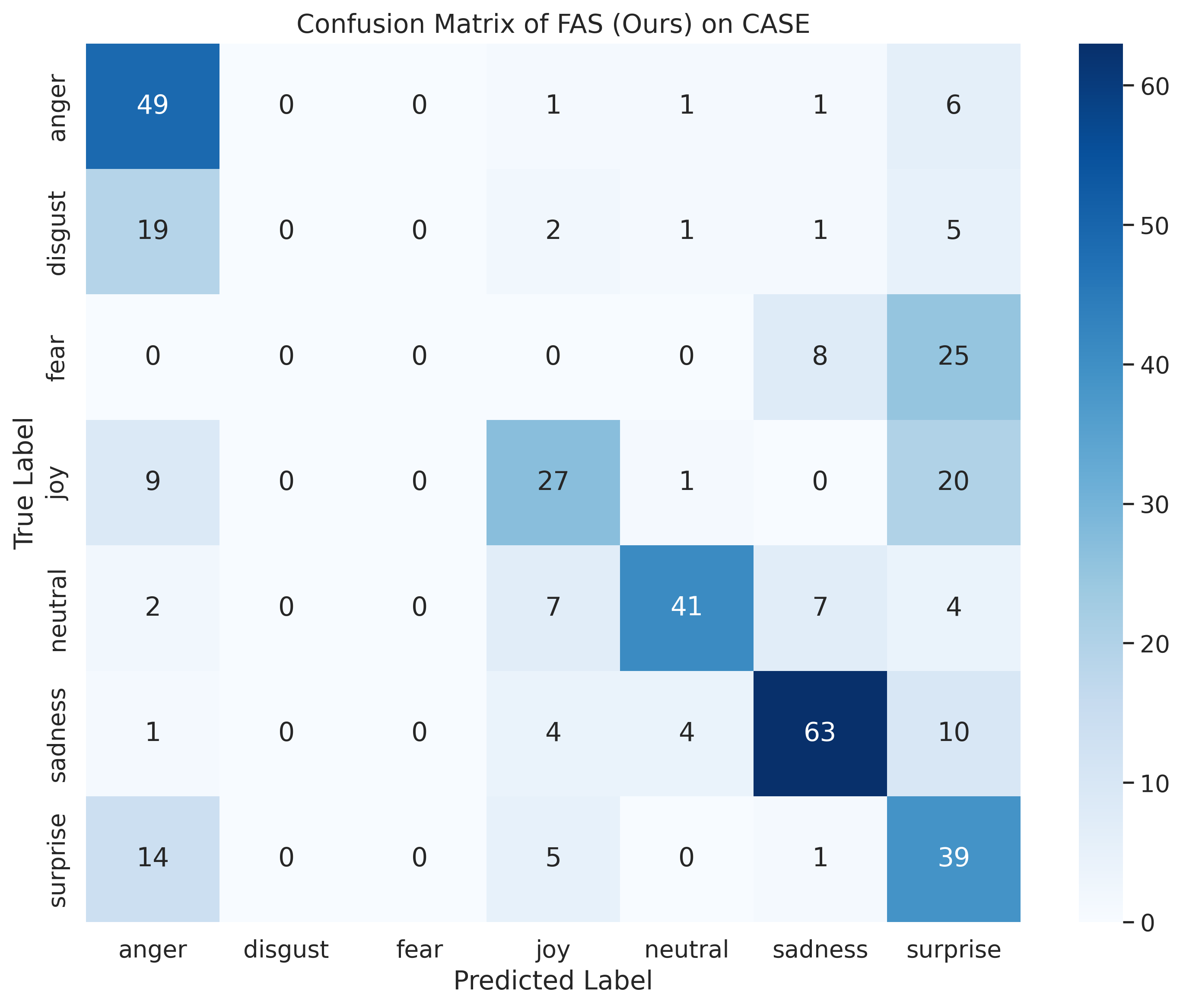}
        \caption{FAS (Ours)}
        \label{fig:CM_CASE_1a}
    \end{subfigure}
    \hfill
    \begin{subfigure}[b]{0.30\textwidth}
        \includegraphics[width=\linewidth]{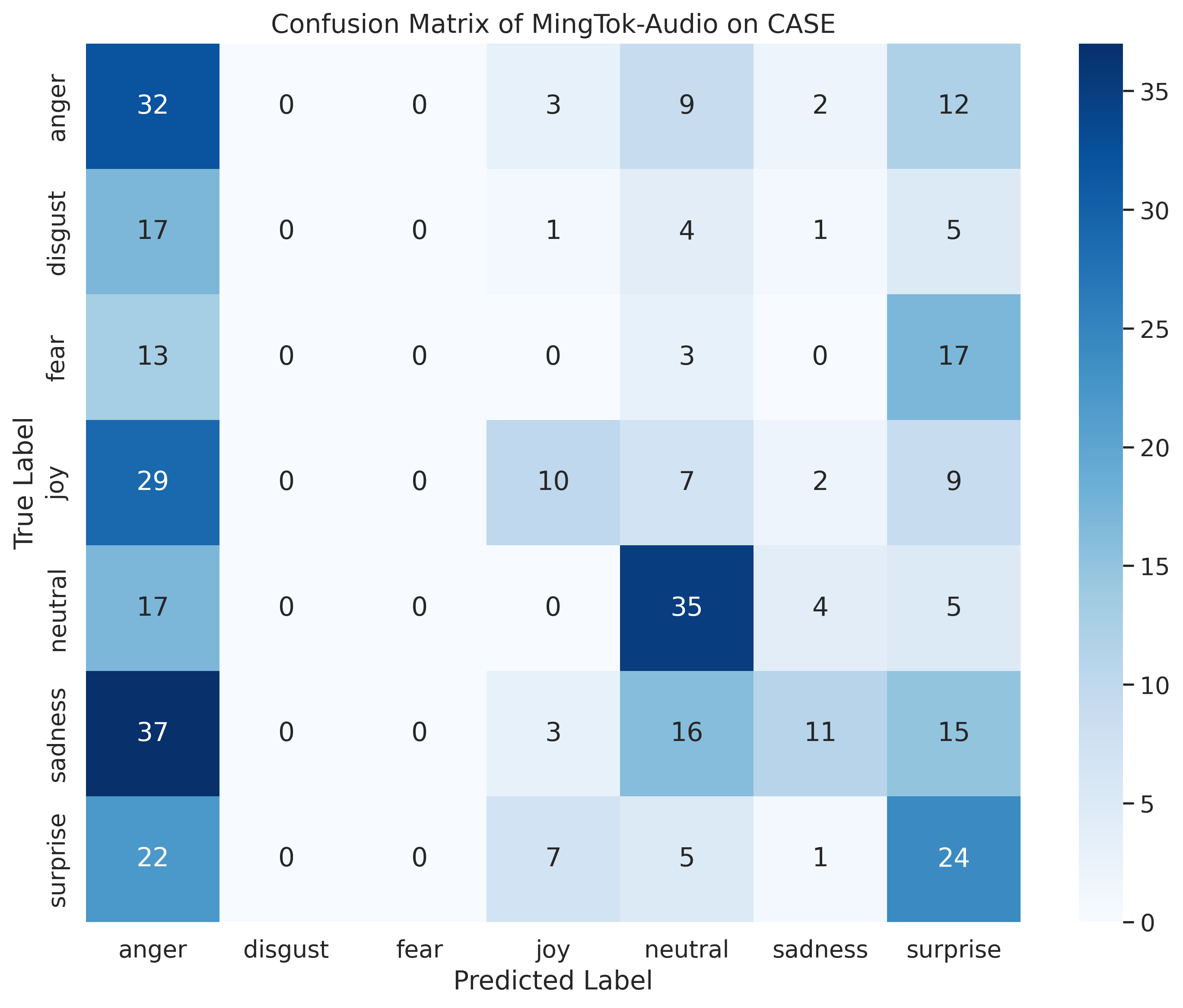}
        \caption{MingTok-Audio \cite{Ming-UniAudio}}
        \label{fig:CM_CASE_1b}
    \end{subfigure}
    \hfill
    \begin{subfigure}[b]{0.30\textwidth}
        \includegraphics[width=\linewidth]{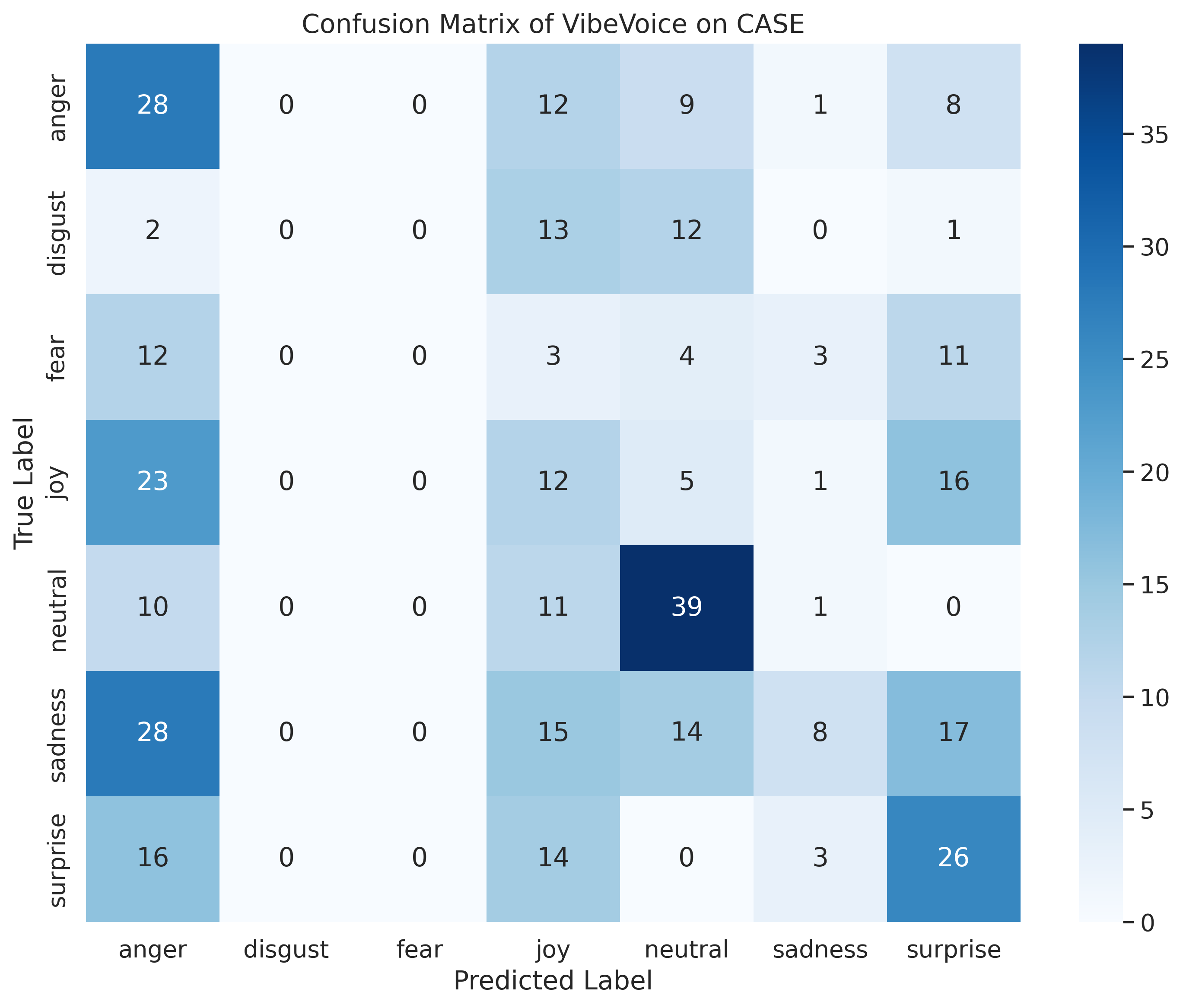}
        \caption{VibeVoice \cite{VibeVoice}}
        \label{fig:CM_CASE_1c}
    \end{subfigure}


    \begin{subfigure}[b]{0.30\textwidth}
        \includegraphics[width=\linewidth]{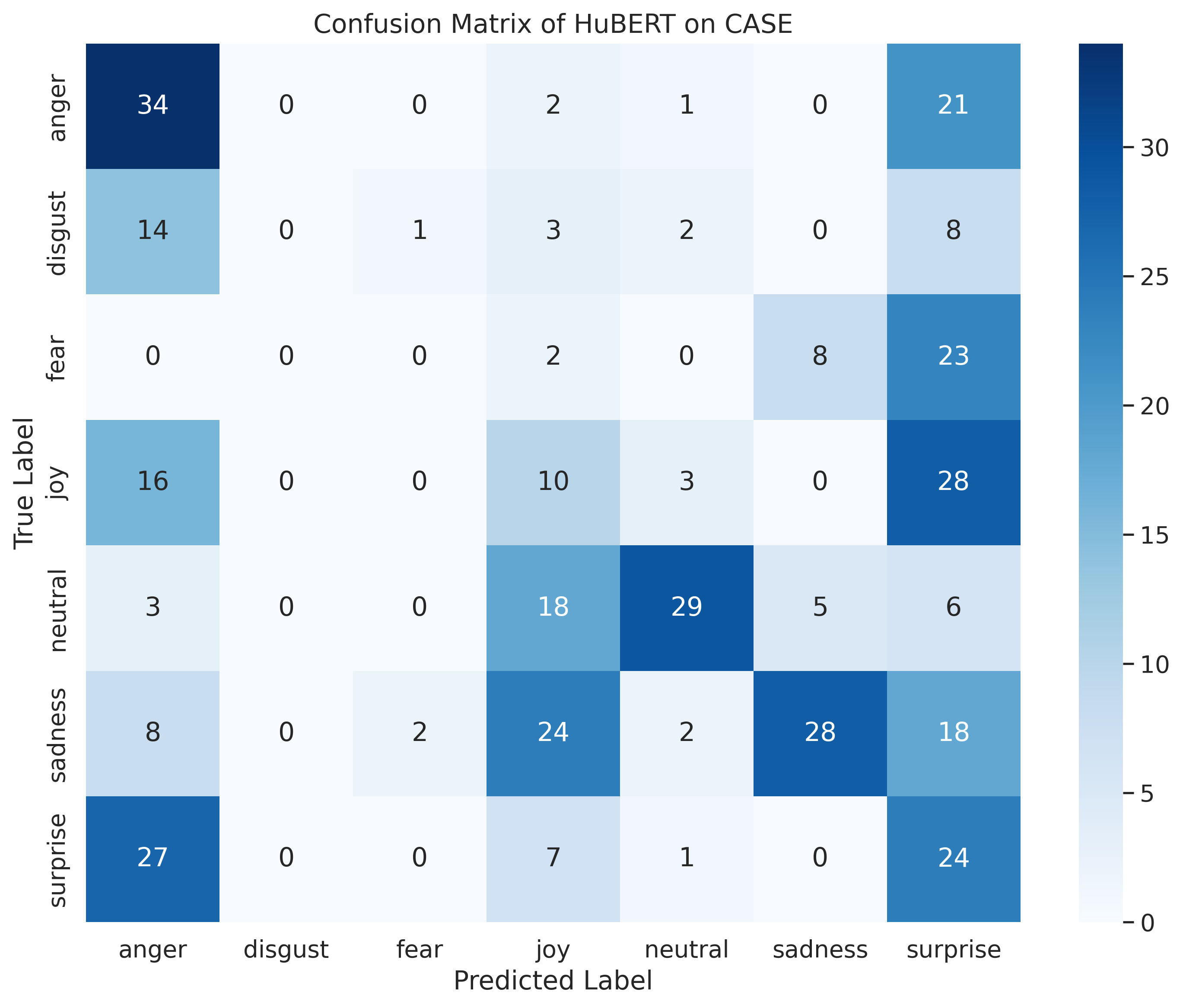}
        \caption{HuBERT \cite{hubert}}
        \label{fig:CM_CASE_2a}
    \end{subfigure}
    \hfill
    \begin{subfigure}[b]{0.30\textwidth}
        \includegraphics[width=\linewidth]{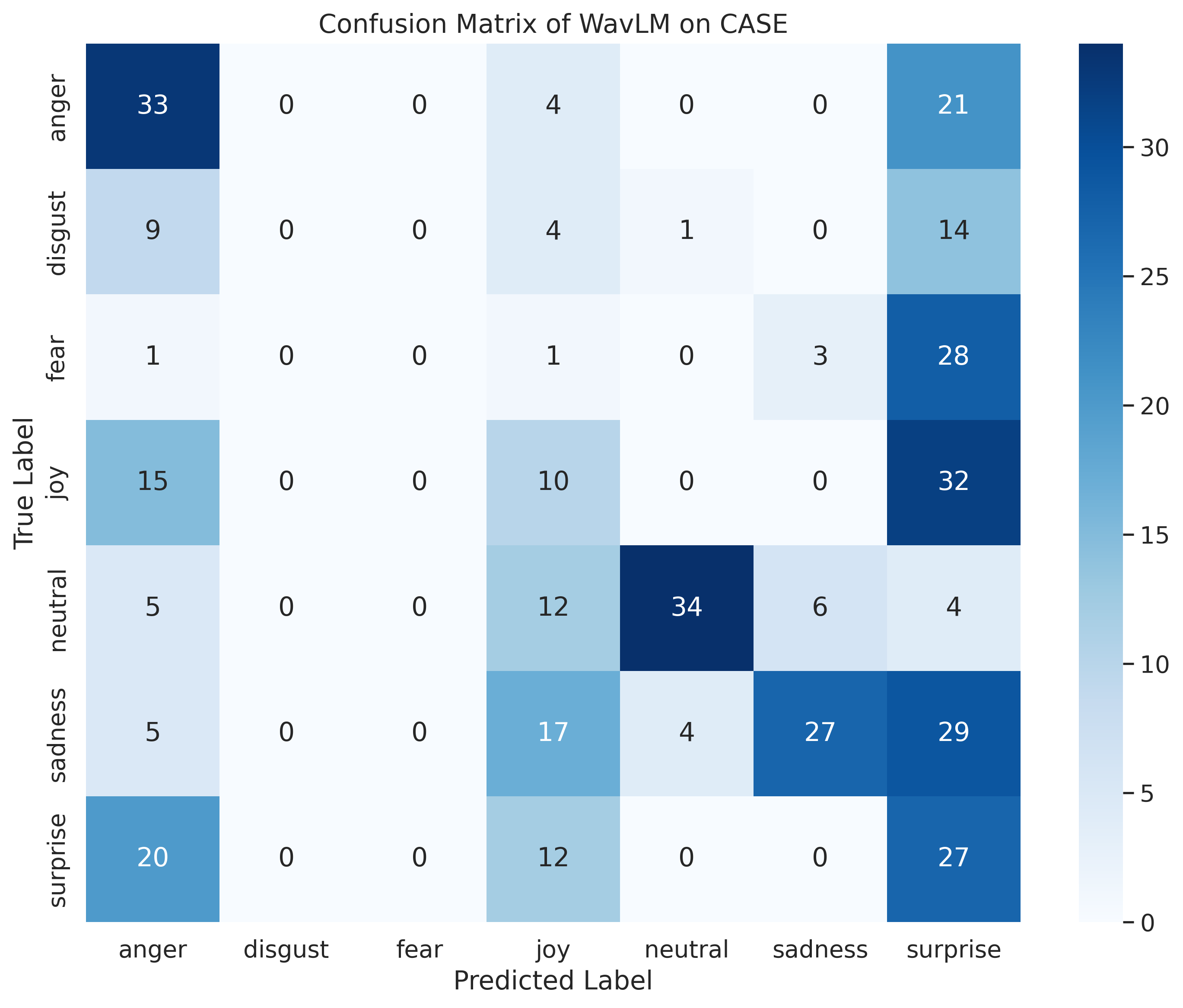}
        \caption{WavLM \cite{Wavlm}}
        \label{fig:CM_CASE_2b}
    \end{subfigure}
    \hfill
    \begin{subfigure}[b]{0.30\textwidth}
        \includegraphics[width=\linewidth]{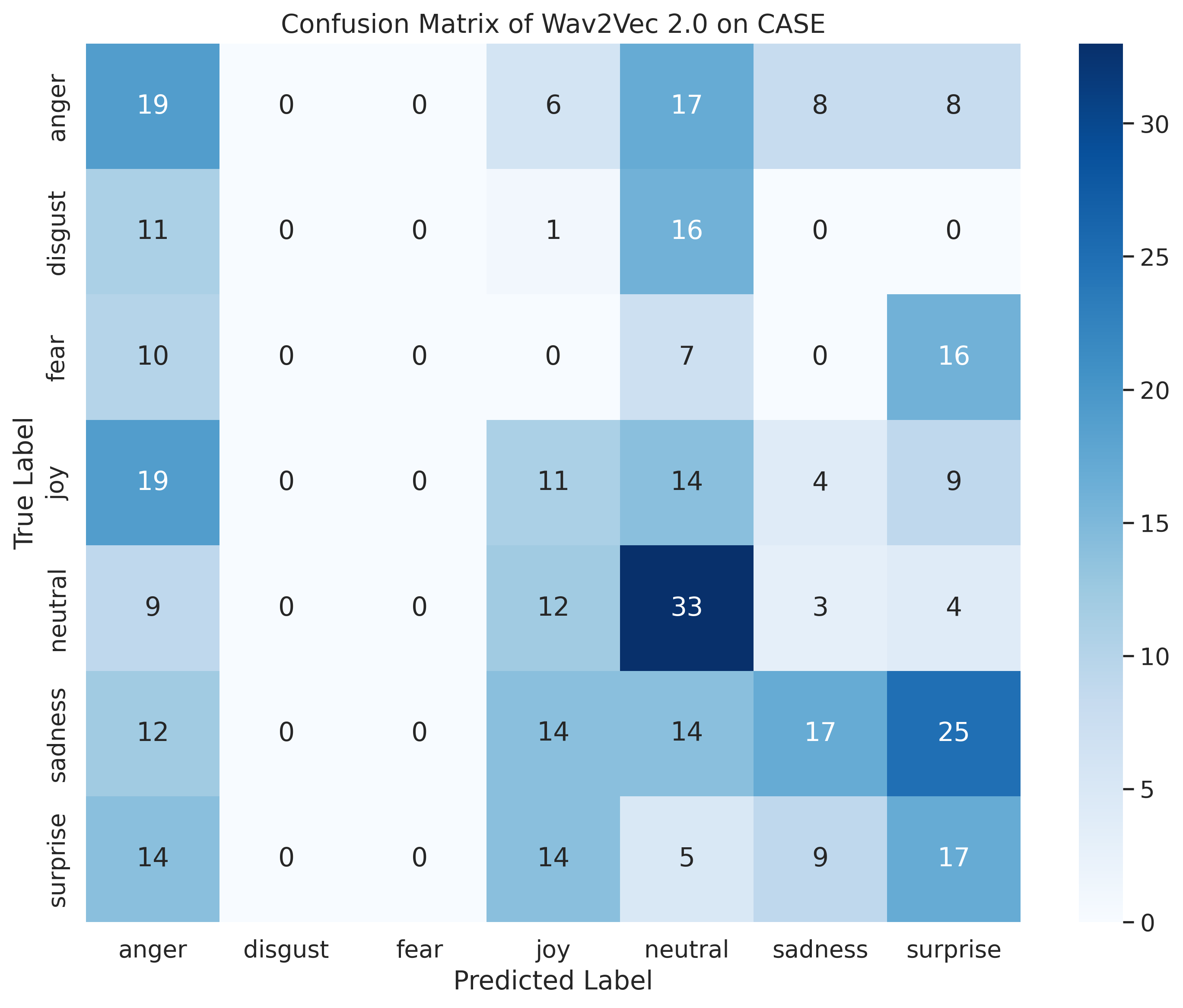}
        \caption{Wav2Vec 2.0 \cite{wav2vec2.0}}
        \label{fig:CM_CASE_2c}
    \end{subfigure}

    \caption{Confusion Matrices on the CASE dataset. Our FAS framework (\ref{fig:CM_CASE_1a}) achieves the highest accuracy on major classes (\textit{anger}, \textit{sadness}, \textit{neutral}) and shows minimal confusion between high-arousal emotions. 
    }
    \label{fig:Confusion_Matrix_CASE}
\end{figure*}

To further evaluate the performance of our proposed FAS framework, we compare its confusion matrix against strong baselines on the CASE and RAVDESS datasets.  
As shown in Figure~\ref{fig:Confusion_Matrix_CASE}, FAS achieves the highest accuracy on major classes—particularly \textit{anger}, \textit{sadness}, and \textit{neutral}—with significantly less confusion between high-arousal emotions (e.g. \textit{anger} vs. \textit{surprise}) than other models.
Notably, no method correctly predicts any samples of \textit{fear} or \textit{disgust}, likely due to the high difficulty of the CASE benchmark and the scarcity of training data that effectively decouples acoustic and semantic cues.

\begin{figure*}[t]
    \centering
    \begin{subfigure}[b]{0.30\textwidth}
        \includegraphics[width=\linewidth]{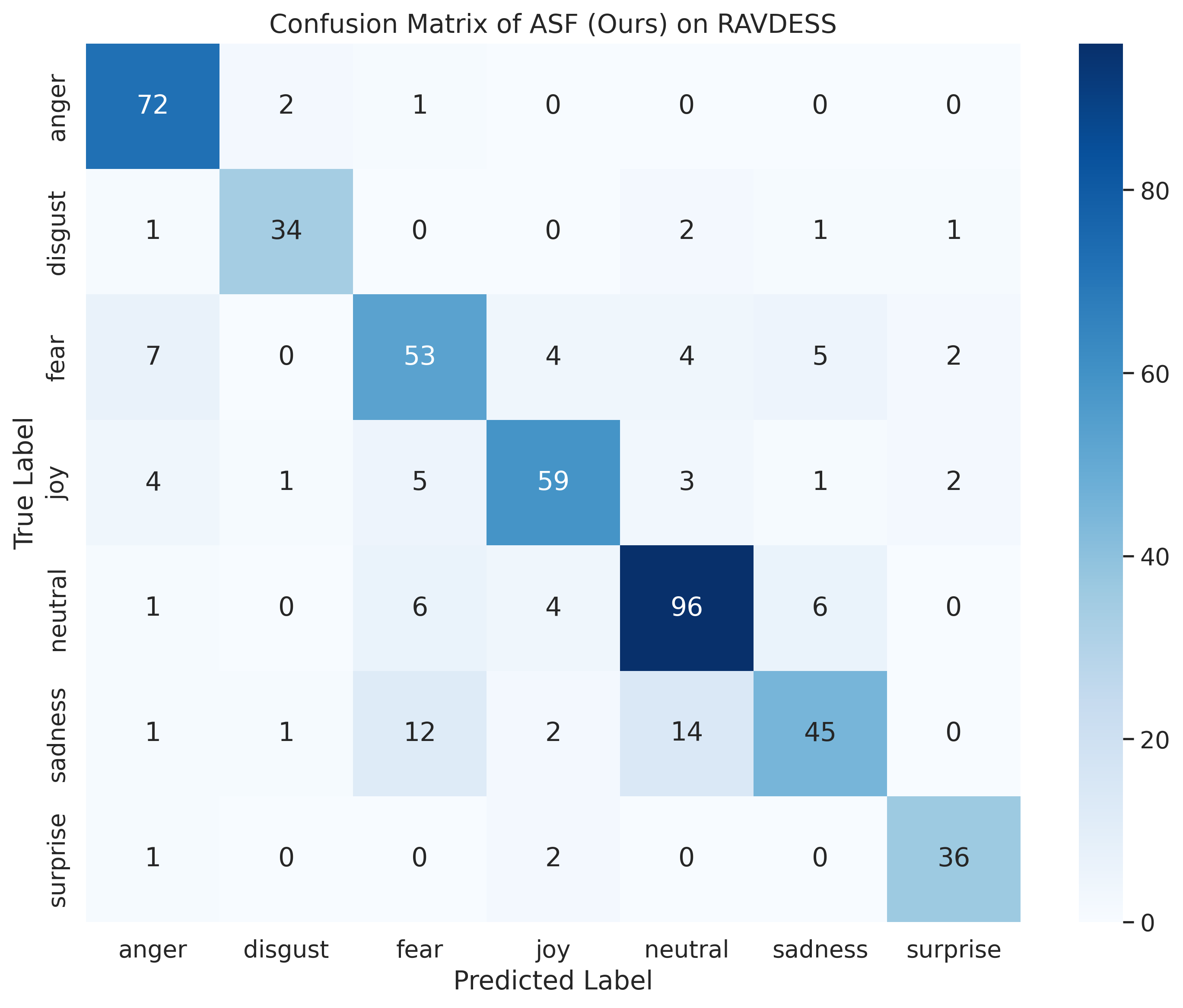}
        \caption{FAS (Ours)}
        \label{fig:CM_RAVDESS_1a}
    \end{subfigure}
    \hfill
    \begin{subfigure}[b]{0.30\textwidth}
        \includegraphics[width=\linewidth]{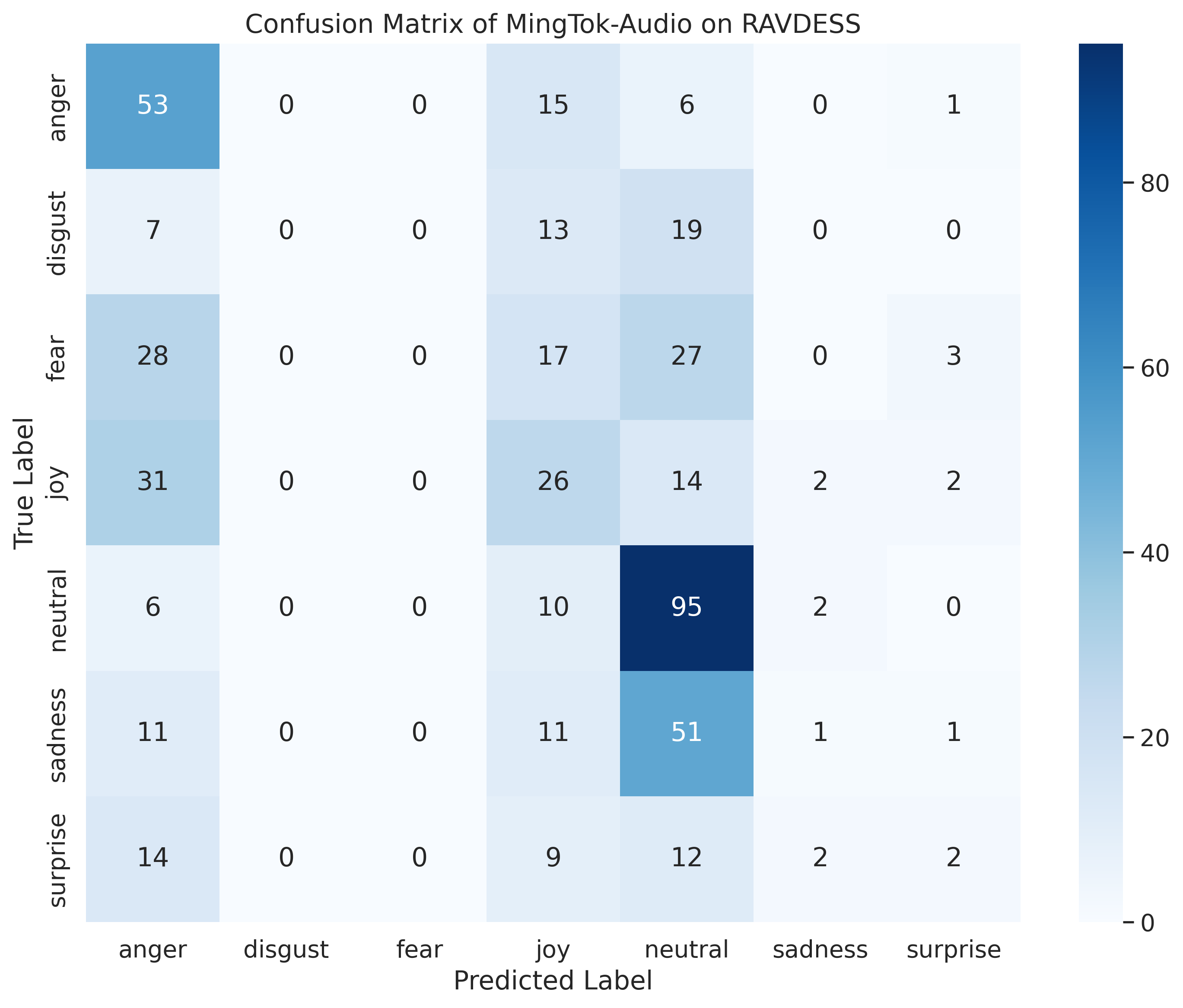}
        \caption{MingTok-Audio}
        \label{fig:CM_RAVDESS_1b}
    \end{subfigure}
    \hfill
    \begin{subfigure}[b]{0.30\textwidth}
        \includegraphics[width=\linewidth]{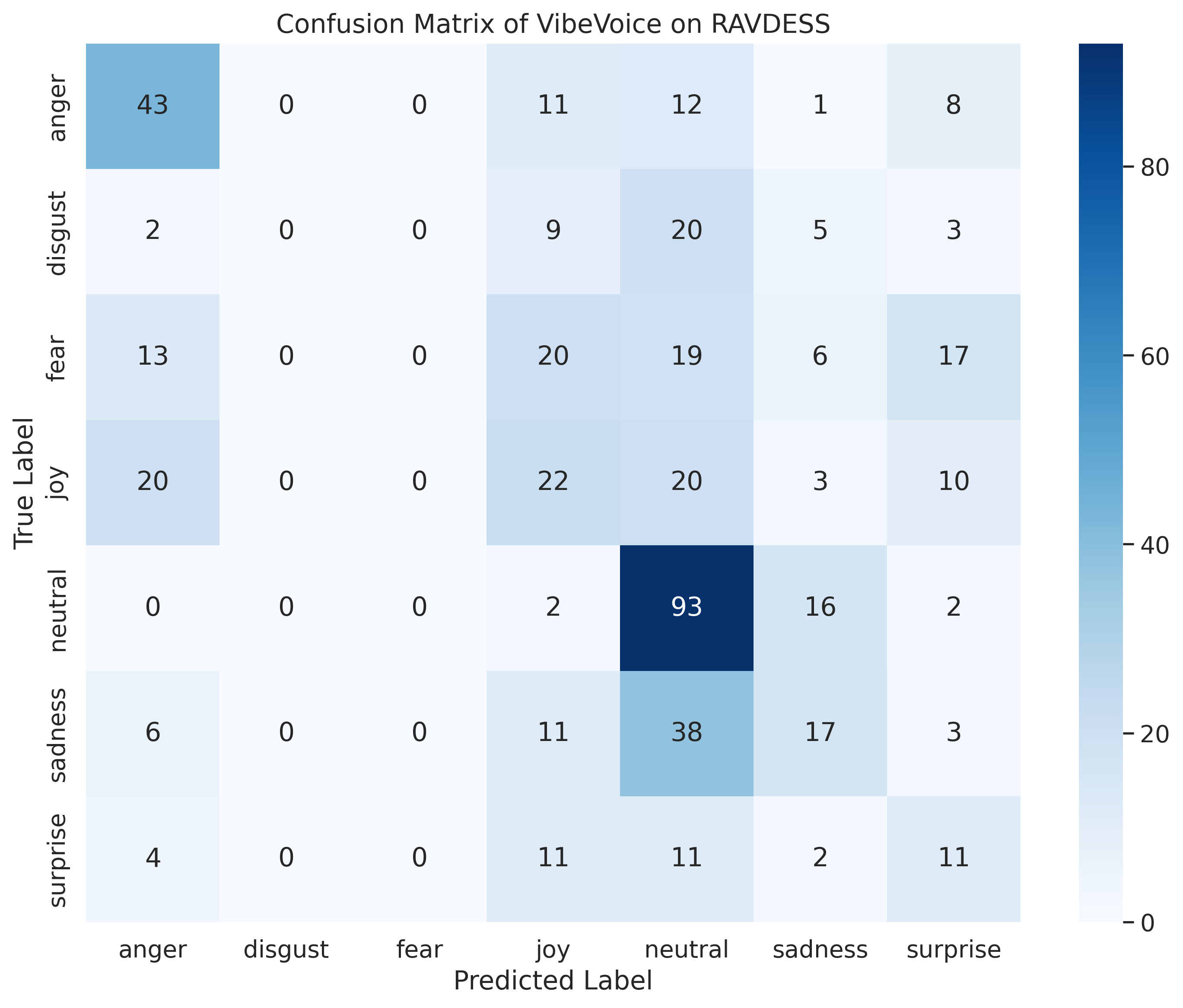}
        \caption{VibeVoice}
        \label{fig:CM_RAVDESS_1c}
    \end{subfigure}


    \begin{subfigure}[b]{0.30\textwidth}
        \includegraphics[width=\linewidth]{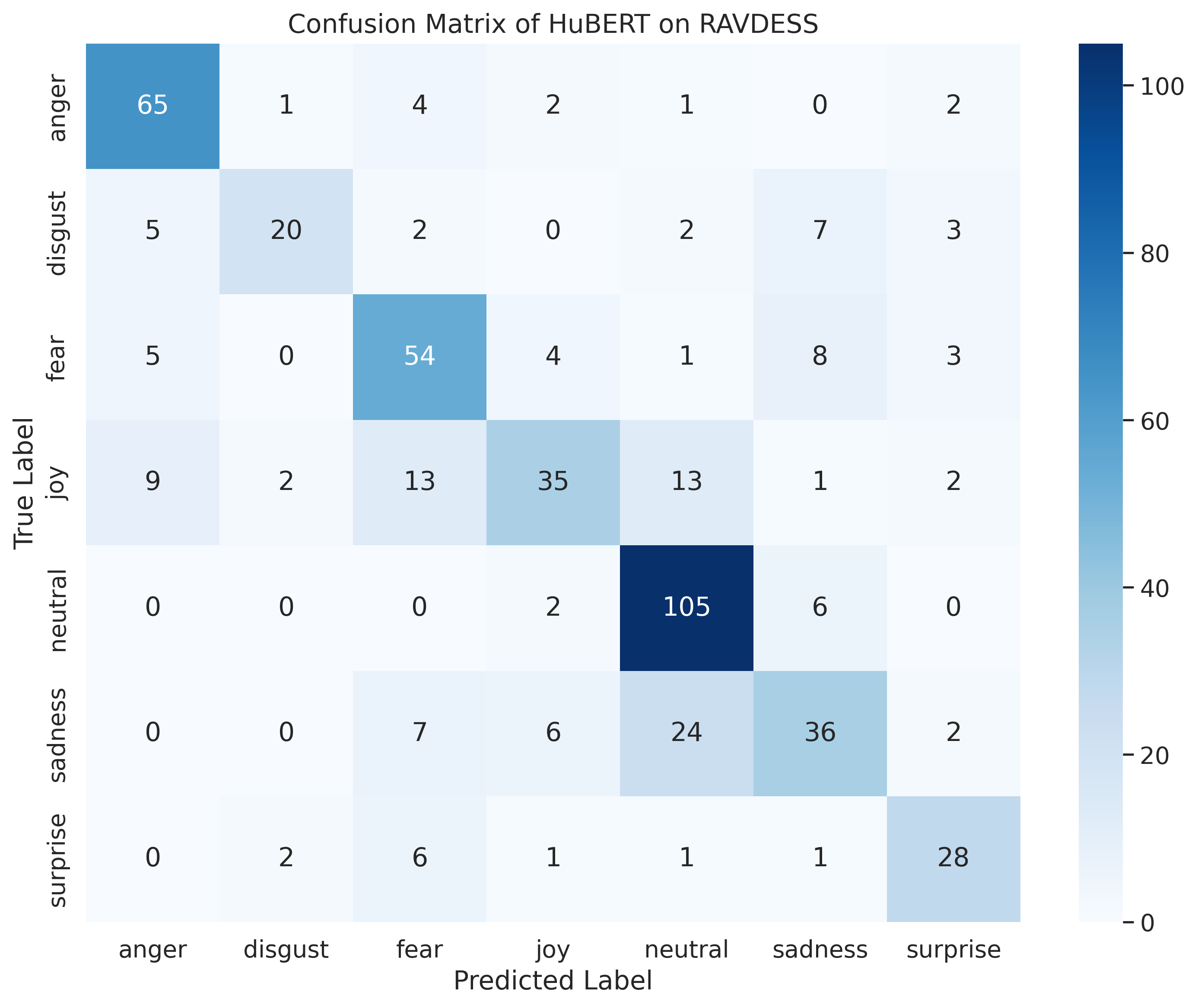}
        \caption{HuBERT}
        \label{fig:CM_RAVDESS_2a}
    \end{subfigure}
    \hfill
    \begin{subfigure}[b]{0.30\textwidth}
        \includegraphics[width=\linewidth]{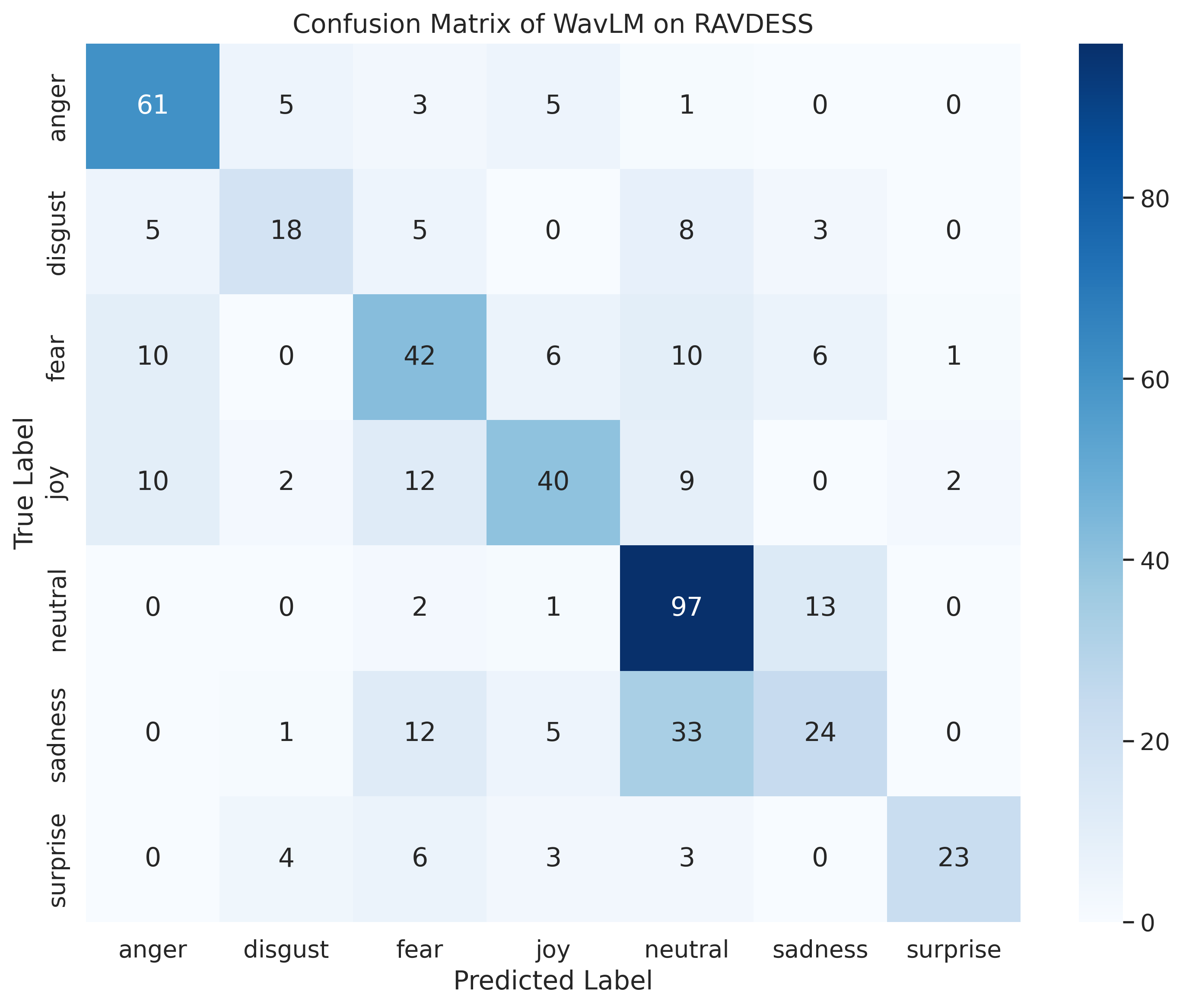}
        \caption{WavLM}
        \label{fig:CM_RAVDESS_2b}
    \end{subfigure}
    \hfill
    \begin{subfigure}[b]{0.30\textwidth}
        \includegraphics[width=\linewidth]{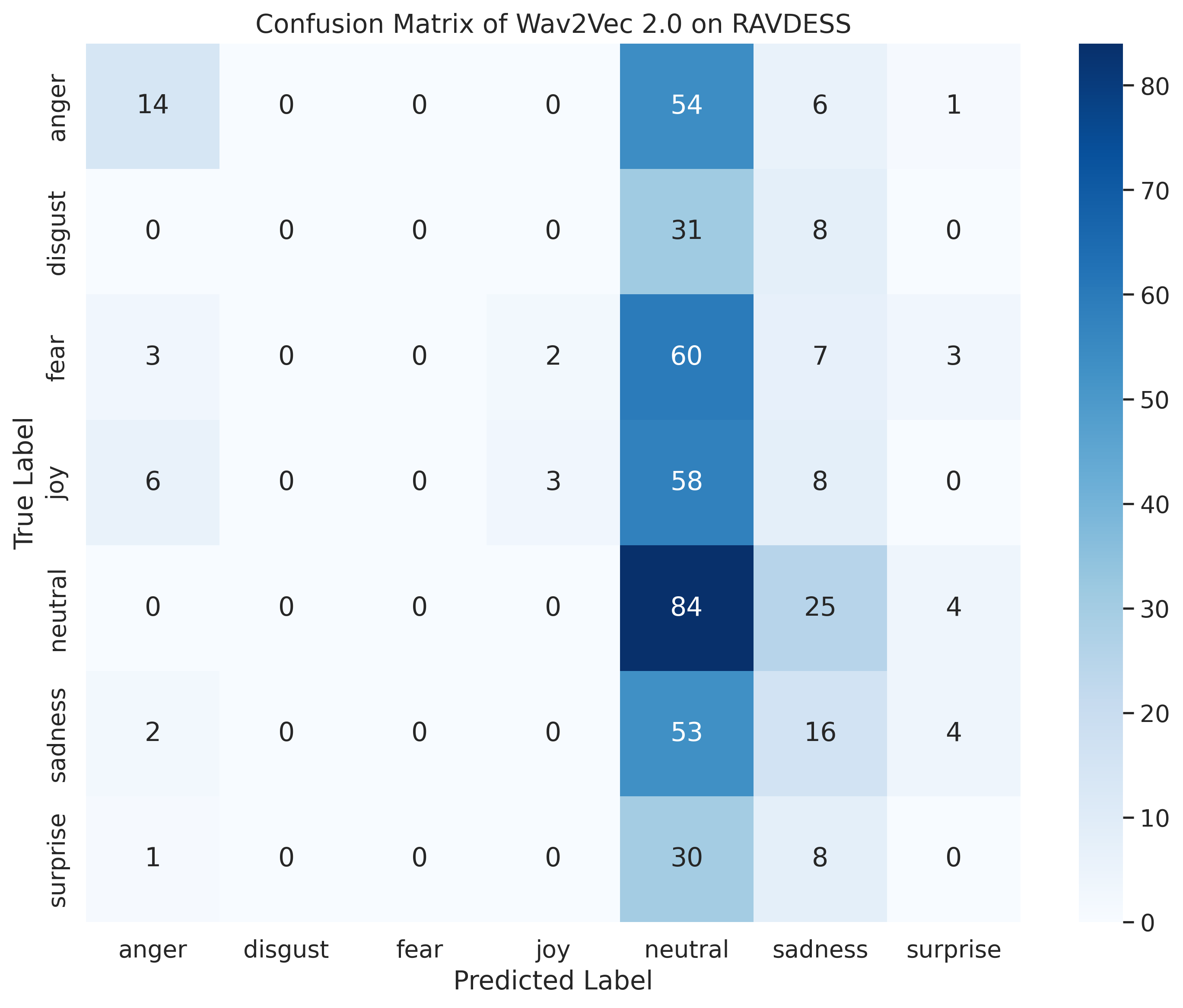}
        \caption{Wav2Vec 2.0}
        \label{fig:CM_RAVDESS_2c}
    \end{subfigure}

    \caption{Confusion Matrices evaluated on the RAVDESS dataset. The FAS framework (\ref{fig:CM_RAVDESS_1a}) demonstrates superior performance across all emotion categories with minimal off-diagonal errors. In comparison, other models including MingTok-Audio (\ref{fig:CM_RAVDESS_1b}), VibeVoice (\ref{fig:CM_RAVDESS_1c}), HuBERT (\ref{fig:CM_RAVDESS_2a}), WavLM (\ref{fig:CM_RAVDESS_2b}), and Wav2Vec 2.0 (\ref{fig:CM_RAVDESS_2c}) show varying degrees of confusion between similar emotions.}
    \label{fig:Confusion_Matrix_RAVDESS}
\end{figure*}

Further validation on the well-structured RAVDESS dataset (Figure~\ref{fig:Confusion_Matrix_RAVDESS}) confirms FAS’s robustness: it exhibits minimal off-diagonal errors and outperforms baselines such as WavLM (\ref{fig:CM_RAVDESS_2b}), VibeVoice (\ref{fig:CM_CASE_1c}), and Wav2Vec 2.0 (\ref{fig:CM_RAVDESS_2c}), especially in distinguishing subtle emotions like \textit{neutral} and \textit{sadness}.

\begin{figure*}[t]
    \centering
    \begin{subfigure}[b]{0.30\textwidth}
        \includegraphics[width=\linewidth]{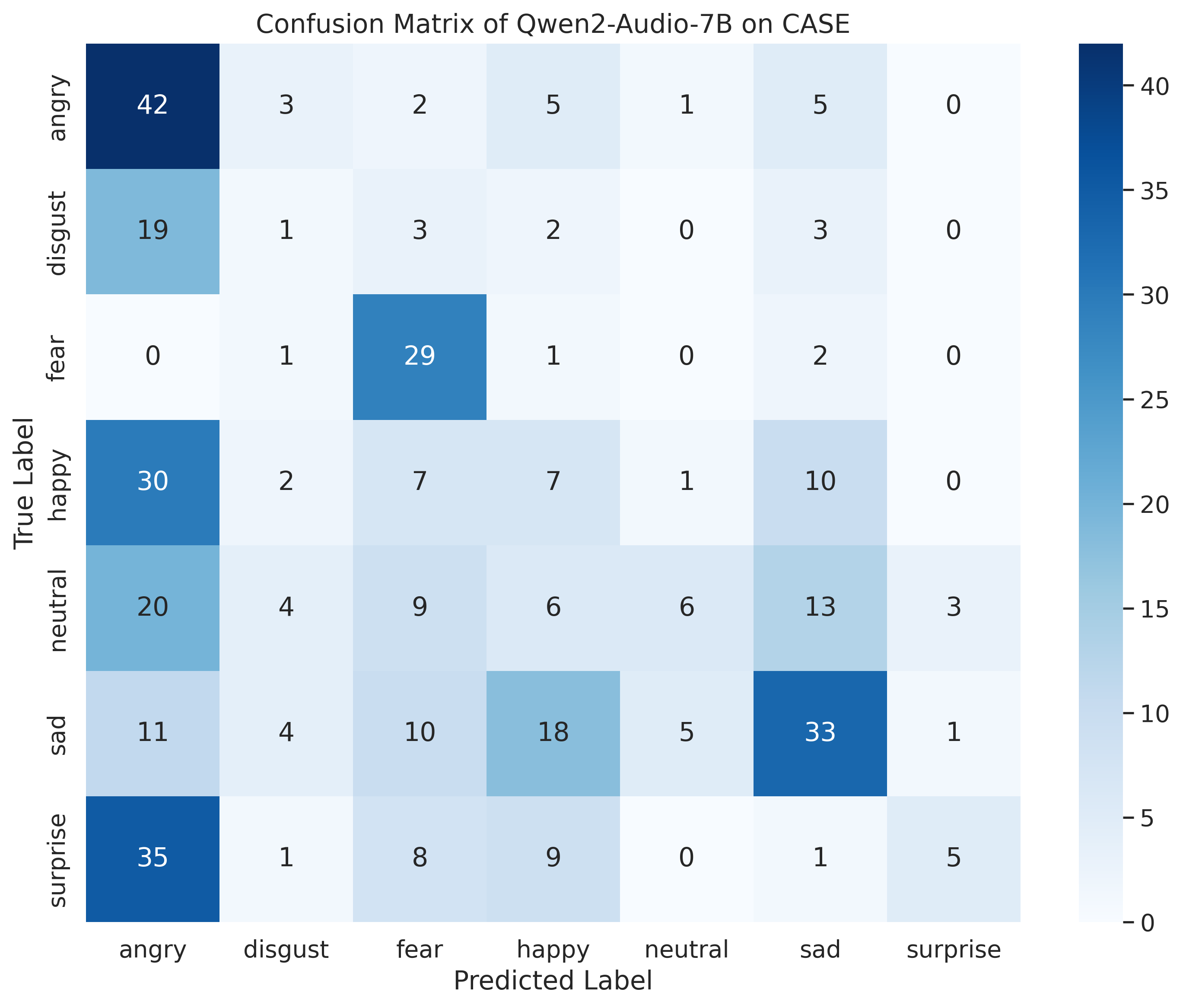}
        \caption{Qwen2-Audio on CASE}
        \label{fig:Qwen2_Audio_CASE}
    \end{subfigure}
    \hfill
    \begin{subfigure}[b]{0.30\textwidth}
        \includegraphics[width=\linewidth]{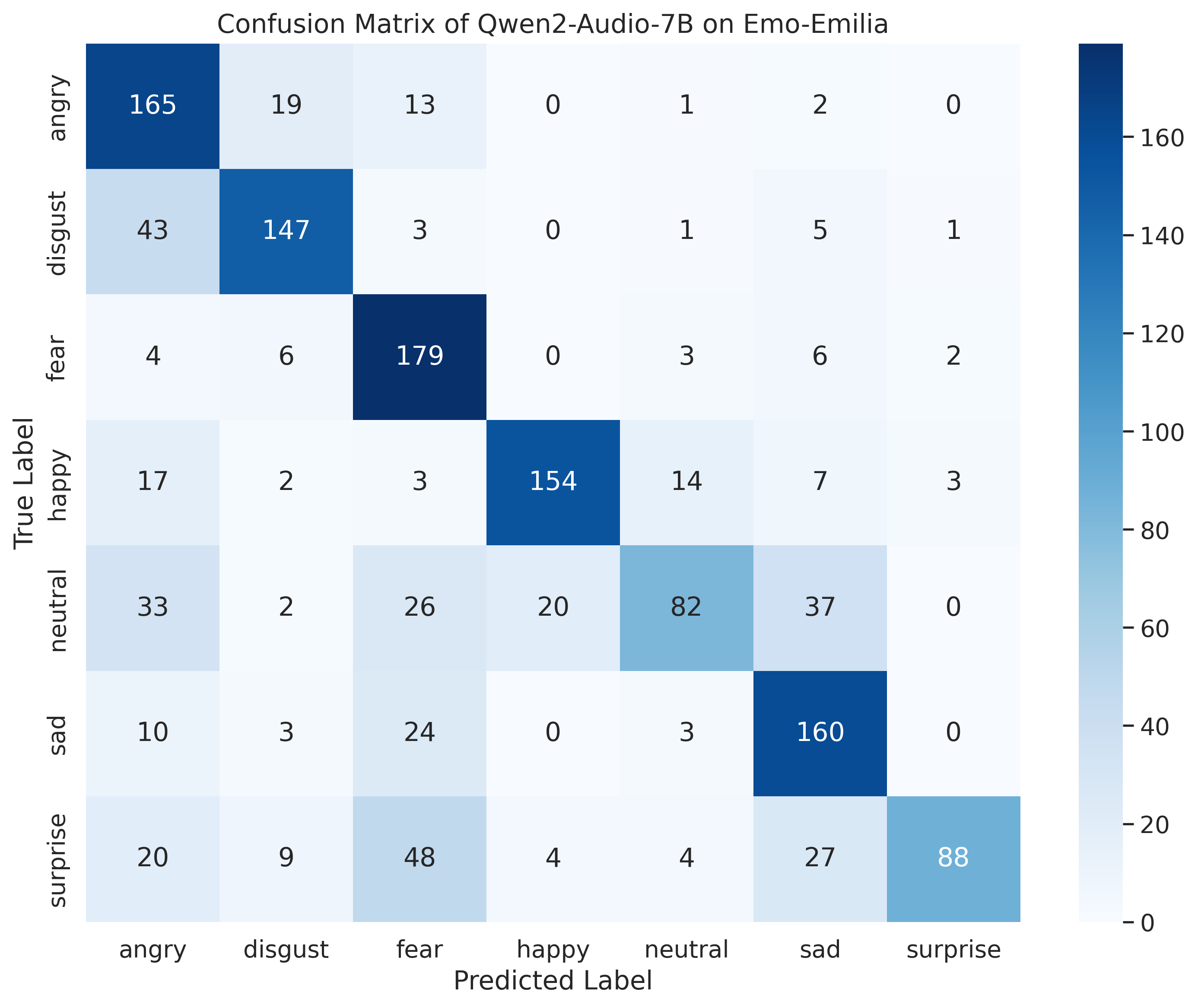}
        \caption{Qwen2-Audio on Emo-Emilia}
        \label{fig:Qwen2_Audio_Emo-Emilia}
    \end{subfigure}
    \hfill
    \begin{subfigure}[b]{0.30\textwidth}
        \includegraphics[width=\linewidth]{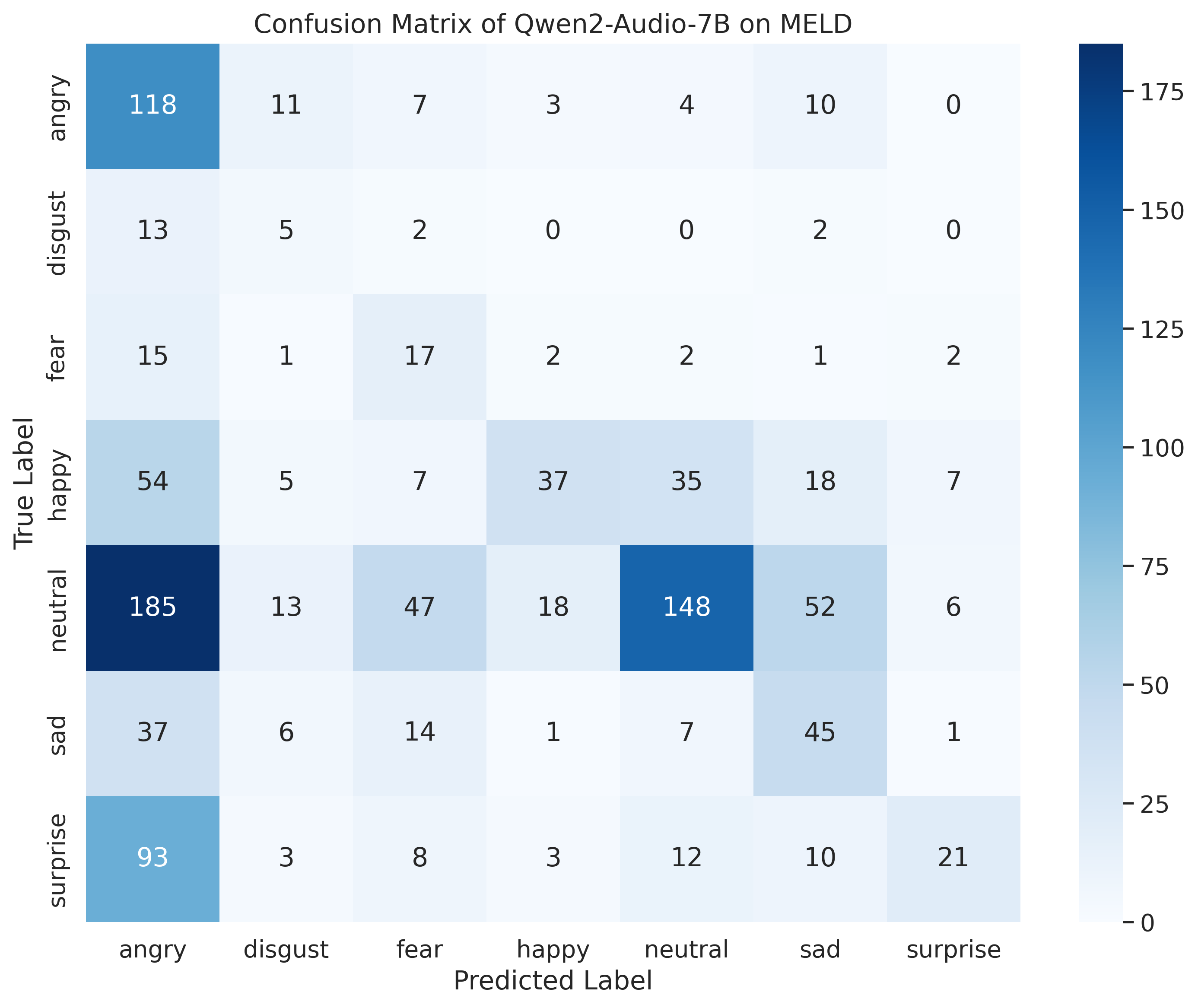}
        \caption{Qwen2-Audio on MELD}
        \label{fig:Qwen2_Audio_MELD}
    \end{subfigure}

    \begin{subfigure}[b]{0.30\textwidth}
        \includegraphics[width=\linewidth]{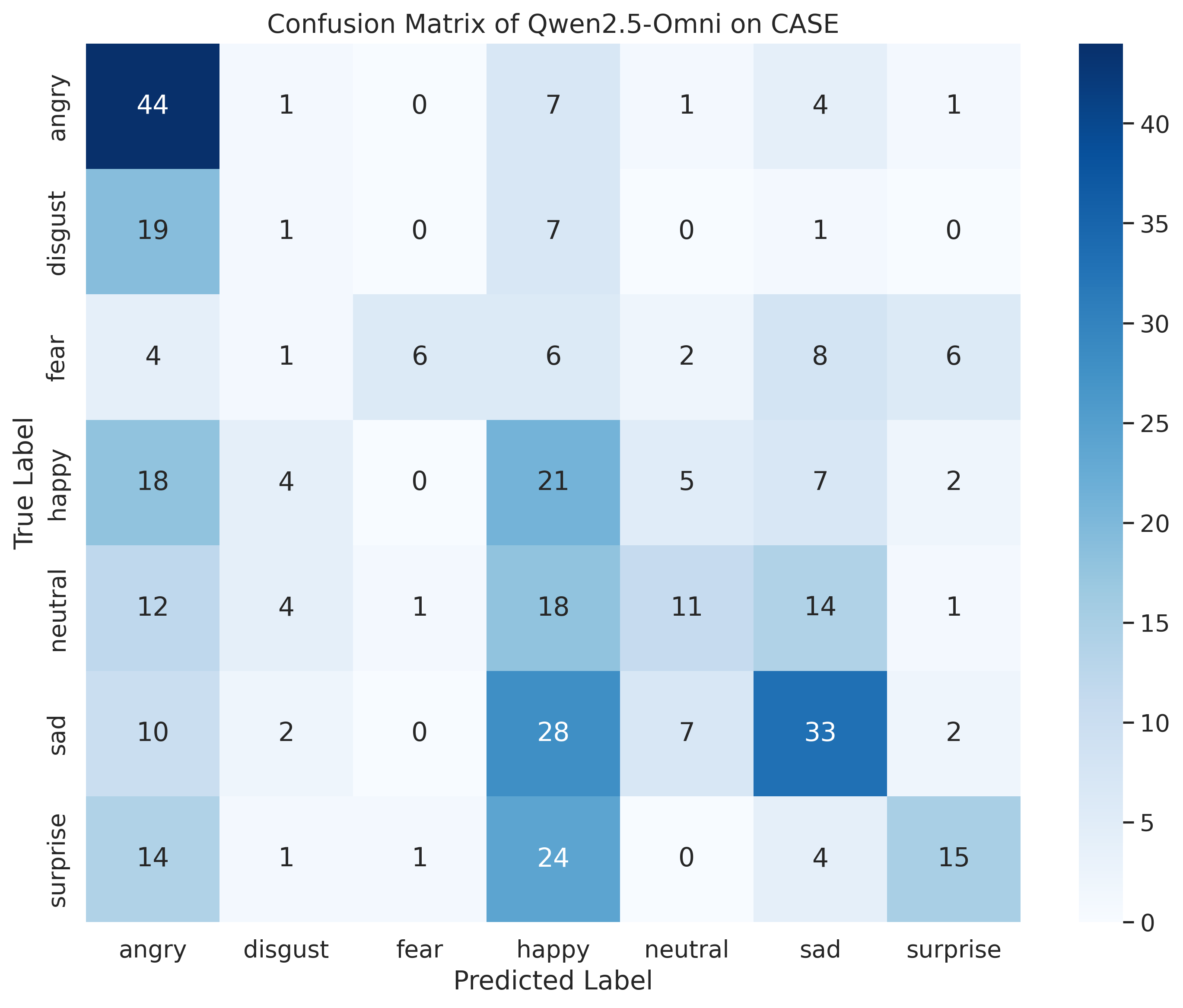}
        \caption{Qwen2.5-Omni on CASE}
        \label{fig:Qwen2.5_Omni_CASE}
    \end{subfigure}
    \hfill
    \begin{subfigure}[b]{0.30\textwidth}
        \includegraphics[width=\linewidth]{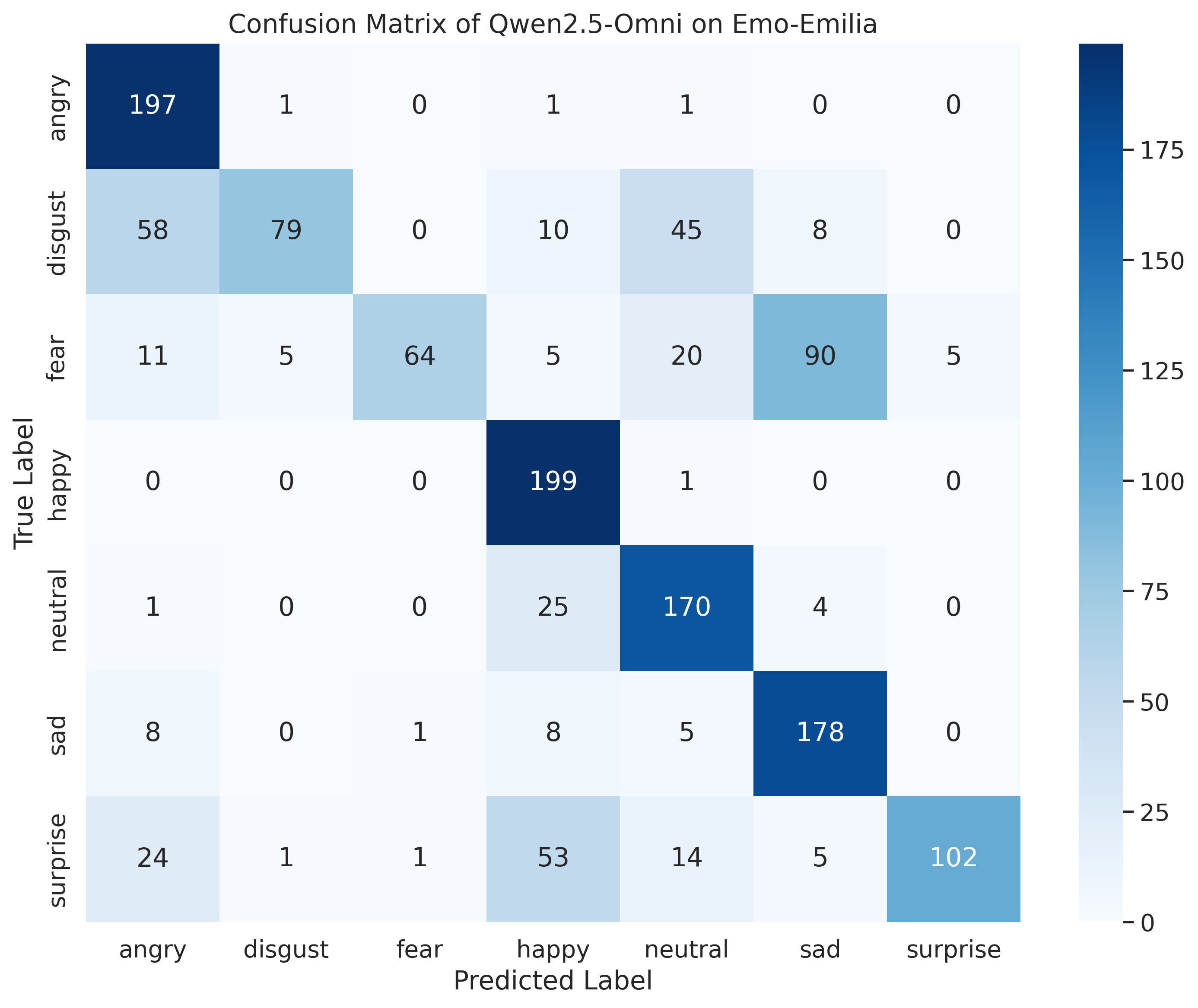}
        \caption{Qwen2.5-Omni on Emo-Emilia}
        \label{fig:Qwen2.5_Omni_Emo-Emilia}
    \end{subfigure}
    \hfill
    \begin{subfigure}[b]{0.30\textwidth}
        \includegraphics[width=\linewidth]{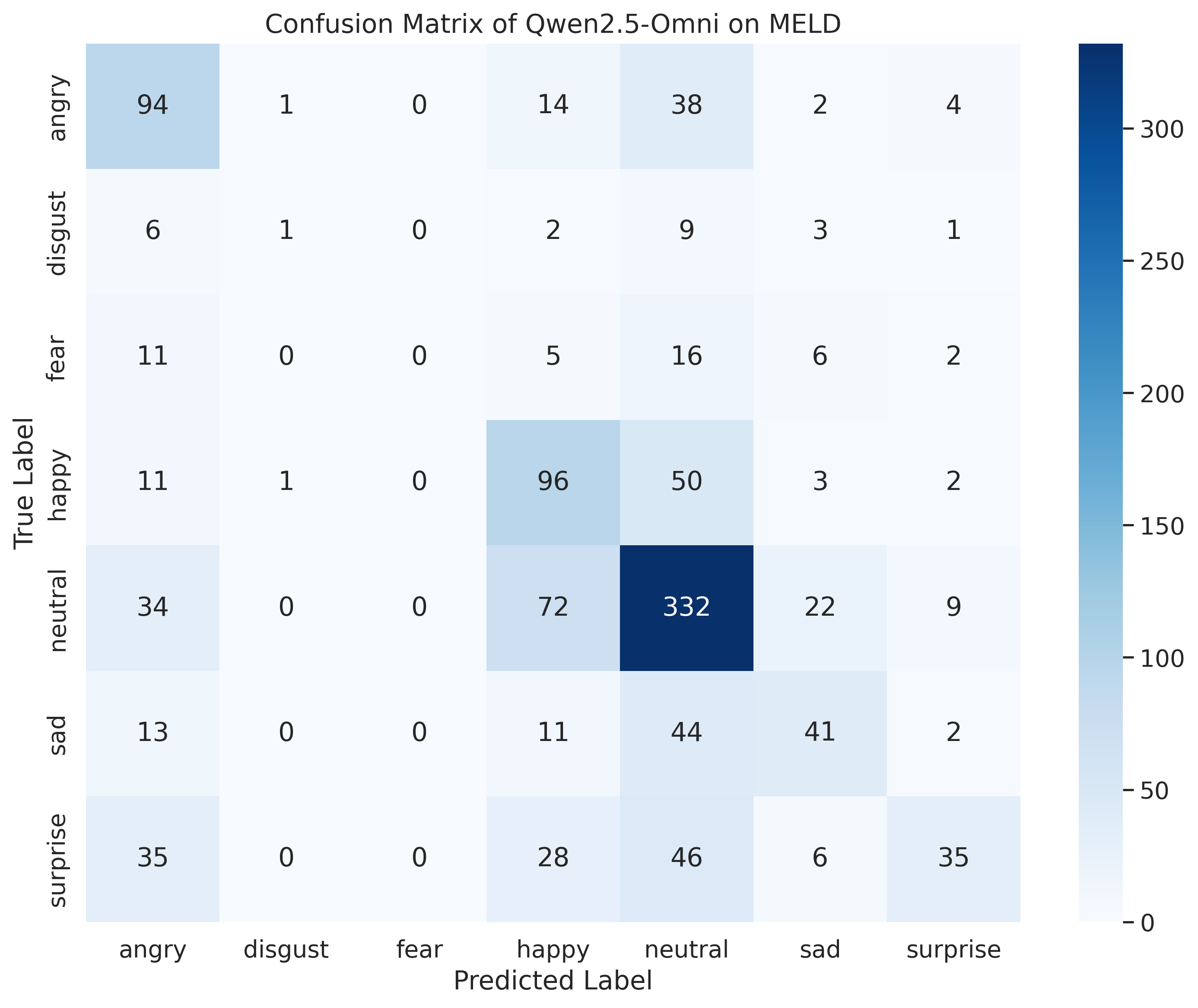}
        \caption{Qwen2.5-Omni on MELD}
        \label{fig:Qwen2.5_Omni_MELD}
    \end{subfigure}

    \caption{Confusion Matrices of Qwen2-Audio-Instrcut and Qwen2.5-Omni on CASE, Emo-Emilia and MELD. }
    \label{fig:Confusion_Matrix_Qwen}
\end{figure*}

In contrast, Audio Language Models (ALMs) show a mixed profile. As illustrated in Figure~\ref{fig:Confusion_Matrix_Qwen}, models like Qwen2-Audio-Instruct and Qwen2.5-Omni achieve reasonable performance on datasets with consistent emotion-expression patterns—such as Emo-Emilia—where their massive pre-training allows them to “memorize” common acoustic-semantic mappings. However, under high-conflict or zero-shot conditions like CASE, they exhibit severe confusion between semantically proximate emotions (e.g. \textit{angry} vs. \textit{disgust}), revealing a fundamental reliance on lexical content over prosodic affect. This brittleness highlights the limitation of implicit fusion in LLM-aligned architectures.

FAS’s consistent accuracy across both structured and ambiguous settings underscores its superior ability to capture fine-grained affective cues by explicitly modeling the interaction—and potential conflict—between acoustic and semantic modalities.

\subsection{Illustrative Samples of CASE Benchmark}
\label{sec:CASE_samples}

To provide concrete examples of the acoustic-semantic conflict embodied in our benchmark, we list a selection of representative samples in Table~\ref{tab:CASE_examples}. Each entry includes the original Chinese utterance, its ground-truth acoustic emotion (what is heard), the inherent semantic emotion (what the words imply), and a short narrative context that justifies the conflict. These scenarios are designed to be psychologically plausible, ensuring that the emotional dissonance arises from realistic human experiences rather than artificial manipulation.

As shown, CASE encompasses a wide variety of psychologically plausible affective conflicts—ranging from fear-laden final words of love spoken by a soldier facing death (Sample 007) to cold contempt disguised as neutral inquiry after an irrational decision (Sample 011). These nuanced scenarios challenge models to disentangle acoustic prosody from semantic content, making CASE a rigorous testbed for evaluating emotional robustness in the presence of cross-modal conflict.

\begin{table*}[htbp]
\centering
\small

\begin{tabularx}{\textwidth}{
    l
    >{\raggedright\arraybackslash}p{5.2cm}
    c
    c
    >{\raggedright\arraybackslash}X
}
\toprule
ID & Utterance & GT Emo. & Semantic Emo. & Contextual Description \\
\midrule
001 & \begin{CJK}{UTF8}{gbsn}那辆卡车失控了，正朝我们冲过来！\end{CJK} & surprised & fear & An adrenaline-seeking extremist expresses morbid excitement at danger. \\
002 & \begin{CJK}{UTF8}{gbsn}你们又赢了，恭喜啊。\end{CJK} & angry & happy & A loser congratulates the winner with barely concealed resentment. \\
003 & \begin{CJK}{UTF8}{gbsn}他走了，再也不会回来了。\end{CJK} & happy & sad & Someone oppressed for years feels secret joy at their tormentor’s departure. \\
004 & \begin{CJK}{UTF8}{gbsn}你给我站住！你到底想怎么样！\end{CJK} & sad & angry & Exhausted from arguing; anger has turned into heartbreak and despair. \\
005 & \begin{CJK}{UTF8}{gbsn}任务完成，目标已清除。\end{CJK} & sad & neutral & A hitman reports completing a mission, but the target was an old friend. \\
006 & Well done—now we’re both trapped. & angry & happy & The speaker sarcastically blames their companion whose reckless actions led to a shared predicament, masking frustration with ironic praise. \\
007 & Mom, Dad... I love you. & fear & happy & A soldier records a final message to his parents before a suicide mission; his voice trembles with terror despite the loving words. \\
008 & No way—he was already dead! & fear & surprised & In a horror scenario, the protagonist witnesses a supposedly slain villain rise again, reacting with visceral fear beneath an initial gasp of shock. \\
009 & The emergency exit is blocked. & fear & neutral & During a fire, someone announces the only escape route is sealed—their tone calm in wording but laced with palpable panic and dread. \\
010 & What? You actually served this? & hate & surprised & A fastidious food lover reacts to a revolting “gourmet” dish with immediate disgust, their shock quickly overtaken by intense loathing. \\
011 & So this is your final decision, then? & hate & neutral & After hearing an utterly unreasonable choice, the speaker delivers a cold, detached confirmation that conveys silent contempt and resignation. \\
012 &By the way, the building is on fire. We should probably leave. & neutral & tension &A character with a dry, British sense of humor and extreme stoicism delivering urgent, life-threatening news in a casual, conversational tone. \\
013 & Oh, a surprise party for me? You shouldn't have. & angry&neutral&An introvert who hates surprises is trying to be polite, but their voice is filled with irritation and anger. \\
014 & I hate you! I never want to see you again! &sad&angry& Saying hateful words during a breakup, but the underlying emotion is one of heartbreak and sadness.\\
015 & I love it. Another spreadsheet.&sad&happy&An employee sarcastically commenting on being assigned more tedious work, their voice full of gloom.\\
016 &You lost the game. It's over.&happy&neutral&A game show host playfully and cheerfully announcing bad news to a contestant. \\
017 & And the winner is... not you. & happy & neutral &A game show host playfully and cheerfully announcing bad news to a contestant.\\
018 &Don't worry about the dishes, I'll just do them. Again.&angry&neutral&A classic passive-aggressive roommate situation. The words are seemingly helpful, but the tone is dripping with anger and resentment. \\
019 &I heard you got the promotion. I am so, so thrilled for you.&sad&excited&Congratulating a coworker who got the promotion they wanted. They are trying to be supportive, but their voice is filled with their own disappointment. \\
020 &You're getting so mad over this little game, it's actually adorable.&happy&angry&A friend playfully teasing and taunting another friend who is getting frustrated while playing a video game.\\

\bottomrule
\end{tabularx}
\caption{Representative acoustic-semantic conflict samples from our proposed CASE benchmark. For full audio demonstrations and additional metadata, please refer to the publicly released dataset files in the open-sourced repository.}
\label{tab:CASE_examples}
\end{table*}

\end{document}